\documentclass[aps,pra,twocolumn,showpacs,groupedaddress,nofootinbib]{revtex4-1}

\usepackage[utf8]{inputenc}  
\usepackage[T1]{fontenc}     
\usepackage[colorlinks,linkcolor=magenta,urlcolor=blue]{hyperref}  
\usepackage{graphicx} 
\usepackage[babel]{microtype}  
\usepackage{amsmath,amssymb,amsthm,bm,mathtools,amsfonts,mathrsfs,bbm} 
\usepackage{xspace}  
\usepackage{centernot}
\usepackage{sistyle}
\usepackage{gensymb}
\usepackage{natbib}
\usepackage{dsfont}
\usepackage{epsfig}
\usepackage{tabularx}
\usepackage{hyperref}

\DeclareMathOperator{\tr}{tr}
\DeclareMathOperator{\Tr}{Tr}

\newcommand{\bra}[1]{\left\langle #1 \right|}
\newcommand{\ket}[1]{\left| #1 \right\rangle}

\newcommand{\ketbra}[2]{\left|#1\middle\rangle\middle\langle#2\right|}

\def\be{\begin{eqnarray}}
\def\ee{\end{eqnarray}}
\mathtoolsset{centercolon}

\newcommand{\figref}[1]{\figurename{~\ref{#1}}}

\usepackage{changes}
\usepackage{comment}

\begin{document}

\title{Demonstration of EPR steering using single-photon path entanglement and displacement-based detection}

\author{T.~Guerreiro,$^1$ F.~Monteiro,$^1$ A.~Martin,$^1$ J.~B.~Brask,$^2$ T.~V\'{e}rtesi,$^3$ B.~Korzh,$^1$ M.~Caloz,$^1$ F.~Bussi\`{e}res,$^1$ , V. B. Verma$^4$, A. E. Lita$^4$, R. P. Mirin$^4$, S. W. Nam$^4$, F. Marsilli$^5$,
M. D. Shaw$^5$, N.~Gisin,$^1$ N.~Brunner,$^2$  H.~Zbinden,$^1$ R.~T.~Thew$^1$}

\affiliation{$^1$Group of Applied Physics, University of Geneva, Switzerland}
\affiliation{$^2$D\'{e}partement de Physique Th\'{e}orique, Université de Gen\`{e}ve, 1211 Gen\`{e}ve, Switzerland}
\affiliation{$^3$Institute for Nuclear Research, Hungarian Academy of Sciences, Debrecen, Hungary}
\affiliation{$^4$National Institute of Standards and Technology, 325 Broadway, Boulder, Colorado, USA}
\affiliation{$^5$Jet Propulsion Laboratory, California Institute of Technology, 4800 Oak Grove Dr., California, USA}

\date{\today}

\begin{abstract}
We demonstrate the violation of an EPR steering inequality developed for single photon path entanglement with displacement-based detection. We use a high-rate source of heralded single-photon path-entangled states, combined with high-efficiency superconducting-based detectors, in a scheme that is free of any post-selection and thus immune to the detection loophole. This result conclusively demonstrates single-photon entanglement in a one-sided device-independent scenario, and opens the way towards implementations of device-independent quantum technologies within the paradigm of path entanglement. 
\end{abstract}

\maketitle

Single-photon entanglement is not only one of the simplest forms of entanglement to generate, it is both fundamentally fascinating and potentially practical. At times its mere existence was debated \cite{Tan1991,*vanEnk2005}, however, today it lies at the heart of key quantum information protocols, such as quantum repeaters~\cite{Sangouard2011}. Path entanglement is generated when a single photon is delocalized over several modes, or paths, e.g. via a 50/50 beam splitter, where it produces a state of the form 
\be \label{singlephot}
\ket{\Psi} =  \frac{1}{ \sqrt{2}} (\ket{0}_A \ket{1}_B + \ket{1}_A \ket{0}_B) \ ,
\ee 
where $A$ and $B$ denote the two entangled output modes. The versatility of this type of entanglement has been demonstrated in experiments for teleportation~\cite{Lombardi2002,*Fuwa2014}, entanglement swapping~\cite{Sciarrino2002,*Osorio2012}, purification~\cite{Salart2010}, the characterisation of multipartite entanglement~\cite{Papp2009,*Grafe2014}, and is the underlying resource for heralded photon amplification~\cite{Ralph2009,*Kocsis2013,*Bruno2016}.

Another direction of interest is to use single-photon entanglement for demonstrations of quantum nonlocality and related device-independent applications. In particular, the combination of single-photon entanglement with weak displacement-based local measurements~\cite{Banaszek1999,Bjork2001} was recently proposed as a practical platform for demonstrating loophole-free Bell-inequality violations~\cite{Brask2012,Brask2013} and device-independent protocols for quantum information processing~\cite{Caprara2015}. Notably, this approach offers a promising alternative to standard setups based on two-photon entanglement, with clear practical advantages, such as the possibility of heralding the entanglement creation (at high rates, e.g. compared to atomic systems \cite{Pironio2010, Hofmann2012, Ritter2012, Hensen2015}) and scalability to networks involving more parties~\cite{Monteiro2015}. 

\begin{figure}
\centering
\includegraphics[width=0.9\linewidth]{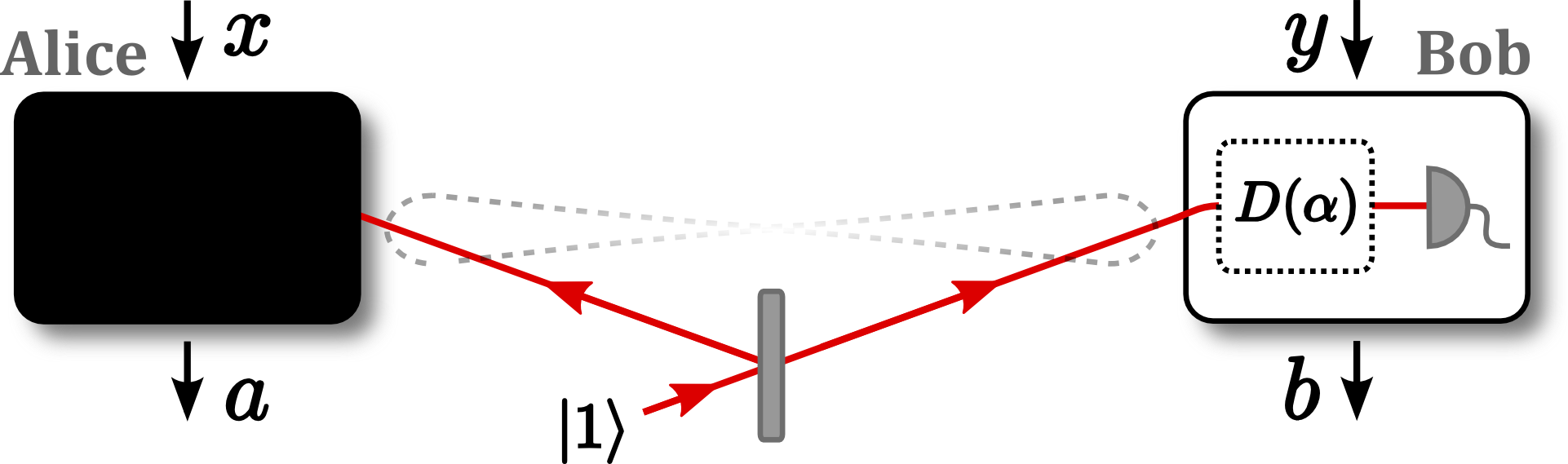} 
\caption{\label{concept} Conceptual view of our steering experiment. Single-photon path entanglement is created by splitting a single photon on a beam splitter. Entanglement between the output modes of the beam splitter is certified in a one-sided DI scenario, via violation of a steering inequality. Alice's device is untrusted (black box), while Bob's device implements characterized (hence trusted) displacement-based detections where the displacement $ D(\alpha) $ is a function of a measurement input~$ y $.}
\end{figure}

Here we report the observation of EPR steering via local weak displacements performed on single-photon entanglement, as sketched in~\figref{concept}. Steering was recently cast in an operational form within quantum information theory~\cite{Wiseman2007}, experimentally demonstrated in Ref.~\cite{Wittmann2012,*Smith2012,*Bennet2012,*Fuwa2015} and represents the key resource for one-sided Device-Independent (DI) protocols, i.e.~where one party is treated as a black-box while the other is trusted~\cite{Branciard2012}. In the following, we first theoretically develop a steering test (a so-called steering inequality~\cite{Cavalcanti2009}) tailored to our setup. We then present an experimental violation of our steering inequality by 4 standard deviations, using a heralded single photon source and an all-fiber displacement-based measurement scheme featuring high-efficiency superconducting nanowire single-photon detectors. As our setup is inherently free of any post-selection, it is immune to the detection loophole~\cite{Brunner2014}. Our experiment thus provides the conclusive demonstration of single-photon path entanglement in a one-sided DI scenario. Moreover, our approach is directly extensible to a loophole-free Bell-inequality test, and thus to the implementation of fully DI protocols~\cite{Acin2007,*Pironio2010}.

{\it Steering.} In a generic steering experiment, as in~\figref{concept}, two separate parties (Alice and Bob) perform local measurements on a shared entangled state. The goal is that Bob can verify that the shared state is indeed entangled without trusting (or equivalently, without any knowledge of) the measurements performed by Alice~\cite{Wiseman2007}. This can be seen as a more stringent test of entanglement than experiments using an entanglement witness, where the measurements of both parties must be well characterised, and less stringent than a Bell inequality test, where none of the parties need to be characterised. 

Formally, Alice receives input $x$ and sends output $a$ to Bob who performs measurements on his system. Specifically, let $\sigma_{a|x}= \tr[\rho (M_{a|x}\otimes \openone )]$ denote the (unnormalized) state of Bob when Alice measures $x$ and obtains outcome $a$, corresponding to some measurement operator $M_{a|x}$. The set of conditional states $\{\sigma_{a|x}\}_{a,x}$ (an assemblage) is termed unsteerable if it can be created by a local strategy without using entanglement, that is, if there exists a local hidden state (LHS) model~\cite{Wiseman2007} compatible with it
\begin{equation}\label{LHS}
\sigma_{a|x} = \sum_\lambda \pi(\lambda) p(a|x\lambda) \sigma_\lambda \hspace{0.5cm} \forall a,x ,
\end{equation}
where $\sigma_\lambda$ represents the LHS, distributed with density $\pi(\lambda)$, and $p(a|x\lambda)$ is Alice's response function. To verify steering, Bob must rule out the existence of an LHS model reproducing the data. This can be certified via violation of so-called steering inequalities~\cite{Cavalcanti2009} (analogous to Bell inequalities).

\begin{figure*}[t]
\centering
\includegraphics[width=0.9\textwidth]{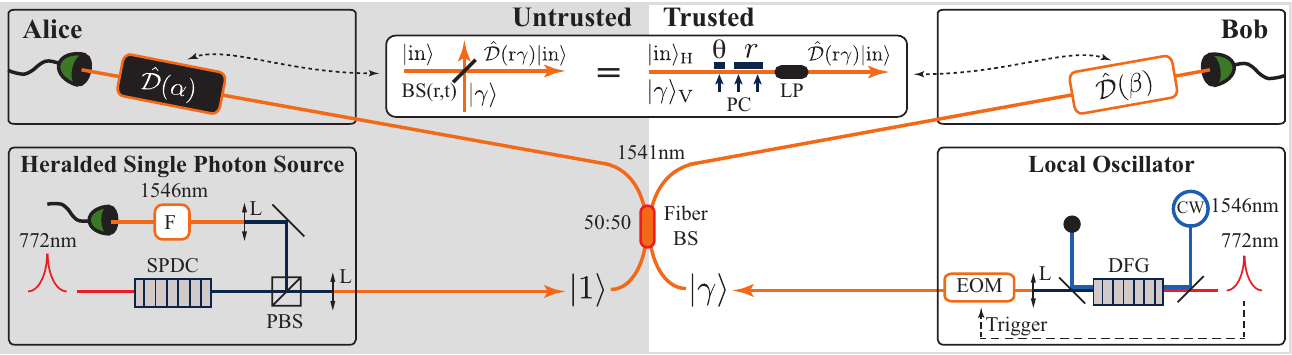}
\caption{\label{setup} Experimental setup. A heralding single photon source is coupled into fiber and incident on a fiber beam splitter (BS), generating heralded entanglement, while local oscillator states are coupled into the same BS with orthogonal polarization. Weak displacements, $\cal{\hat{D}(\alpha)}$, $\cal{\hat{D}(\beta)}$, are performed in an all-fibre configuration (inset) followed by single photon detectors that constitute the displacement-based detection. See main text for details and notation.}
\end{figure*}

{\it Steering inequality for displacements.} In order to demonstrate steering of single-photon entanglement, we must first construct a steering inequality tailored to the displacement-based measurements. The latter consist in an optical displacement $ D(\alpha) $ followed by (non-photon-number-resolving) single-photon detections. In practice they can be implemented by mixing the mode to be measured with a local oscillator (LO) on a highly transmissive beam splitter, and then detecting the transmitted mode, while the reflected mode is discarded~\cite{Paris1996} (see inset \figref{setup}). A no-click outcome corresponds to the projector $\Pi(\alpha) = \ket{\alpha}\bra{\alpha}$, where $\ket{\alpha}$ is a coherent state corresponding to the displacement $\alpha =  r e^{i\theta}$, where $r \geq 0$ and $ \theta \in [0 , 2\pi ]$~\cite{Banaszek1999,Brask2012}. Assigning outcomes $\pm 1$ to the click/no-click events respectively, a displacement measurement then corresponds to the observable $M(\alpha) = 2\ketbra{\alpha}{\alpha} - \openone$. Note that such measurements are always conclusive, as no-click events are not discarded. 

Deriving a steering inequality for our setup is nontrivial because Bob's measurement operator, with binary outcomes, lives in an infinite dimensional Hilbert space. However, we can take advantage of the fact that our target state (of the form \eqref{singlephot}) lives in the 0-1 photon subspace, i.e. a simple qubit subspace. We thus first derive a steering inequality valid in the qubit subspace (using existing methods developed for discrete systems \cite{Pusey2013,Skrzypczyk2014}) and then extend it to the full space.

Specifically, we derive a steering inequality for a scenario with four measurements for Alice ($x=1,2,3,4$) and binary output $a=\pm 1$ of the form 
\begin{align}
\label{eq.ineq01}
S' = \tr \left[ G_R'\sigma_R' + \sum_{x=1}^{4} G_x' \sigma_{+|x}' \right] \leq S_{max}' ,
\end{align}
where $\sigma_R' = \sigma_{+|x}' + \sigma_{-|x}'$ is the reduced state of Bob, $G_R'$ and $G_x'$ are $2 \times 2$ matrices (see Appendix for details). The inequality holds for any unsteerable assemblage (i.e.~admitting a decomposition of the form \eqref{LHS}), hence violation of the inequality certifies steering. The bound $S_{max}'$ is given by the largest eigenvalue of the matrices $G_R' + \sum_x l_x G_x'$, considering any possible deterministic strategy labelled by $l_x=0,1$ (see Appendix).

Next, we consider the restriction of $\Pi(\alpha)$ to the qubit subspace
\begin{equation}
\label{eq.M01}
\Pi'(\alpha) = \left( \begin{array}{cc}
e^{-r^2} & e^{-r^2-i\theta} r \\
e^{-r^2+i\theta} r & e^{-r^2}r^2 \end{array} \right) .
\end{equation} 
By choosing a set of amplitudes $\alpha_y$, we can get a set of operators that span the $2 \times 2$ space together with the identity. The $G'$ matrices can then be resolved on these (note that the decomposition is not necessarily unique when the operators $\Pi'(\alpha_y)$ are not linearly independent)
\begin{equation}
\label{eq.Gdecomp}
G_\nu' = \sum_{y=1}^4 c_{\nu y} \, \Pi'(\alpha_y) + c_{\nu 0} \mathbbm{1} ,
\end{equation}
for some real coefficients $c_{\nu y}$. For our experiment, we take four settings on Bob side (labeled by $y=1,...,4$), given by amplitudes $\alpha_y$, with fixed $r>0$ and phases $\theta \in \{0,\pi/2,\pi,3\pi/2\}$. The measurement outcome is denoted $b=\pm 1$. We now construct an expression analogous to \eqref{eq.ineq01} in the full space as follows. We define
\begin{equation}
S = \tr \left[ G_R\sigma_R + \sum_{x=1}^{4} G_x \sigma_{+|x} \right] ,
\end{equation}
with 
\begin{equation}
\label{eq.Gdecompfull}
G_\nu = \sum_{y=1}^4 c_{\nu y} \, \Pi(\alpha_y) + c_{\nu 0} \mathbbm{1},
\end{equation}
where we are no longer restricted to the 0-1 photon subspace. Similarly to $S'$, the quantity $S$ defines a steering inequality with the bound given by the maximal eigenvalue of the matrices $G_R + \sum_x l_x G_x$ (where as before $l_x=0,1$). This value can be found approximately by introducing a cut-off in photon number. If $r$ is not too large then one can check that the cut-off need not be very high (e.g.~for $r=0.2$, a maximal photon number of 4 is sufficient and the contribution of 5 photons or more is negligible). Thus, we arrive at a steering inequality $S \leq S_{max}$. In general, the bound will be larger than in the subspace, i.e. $S_{max} > S_{max}'$.

The expression $S$ can be computed directly from the experimental data. Using \eqref{eq.Gdecompfull} and the definitions of $\sigma_R$ and $\sigma_{+|x}$ we can rewrite $S$ in terms of the observed conditional probabilities $p(a,b|x,y)$, and obtain the steering inequality
\begin{equation}
\label{eq.ineqprobs}
S = \sum_{x,y=1}^4\sum_{a,b = \pm 1} c^{ab}_{xy}\, p(a,b|x,y) + c_0 \leq S_{max},
\end{equation}
for a new set of real coefficients $c^{ab}_{xy}$, $c_0$ (see Appendix). Numerical optimization shows that the violation of the above steering inequality is possible using a single-photon entangled state \eqref{singlephot}, provided the total transmission and detection efficiency is above $\sim 43 \%$.

{\it Experiment.} The experimental setup is shown in~\figref{setup}. The heralded single photon source (HSPS) is based on Type-II spontaneous parametric down-conversion (SPDC) in a PPKTP crystal satisfying the phase-matching condition  \SI{772}{nm} $\rightarrow $ \SI{1541}{nm} $+$ \SI{1546}{nm}. The HSPS is pumped by a Ti:Saphire laser in picosecond regime to generate pure ($>$~90\%) photons without frequency filters~\cite{Bruno2014}. The probability of generating a photon pair was set to $ 10^{-3} $. The photons are then separated by a polarizing beam splitter (PBS). Detection of one photon at \SI{1546}{nm} heralds the presence of a single photon at \SI{1541}{nm} in the mode of an optical fibre with a heralding efficiency close to 80\%\footnote{In principle the heralding efficiency can be further increased to >90\% with the use of high transmission optical components \cite{Guerreiro2013}.}. A \SI{0.5}{nm} interference filter is placed on the heralding photon path of the HSPS so that the purity of the heralded photon approaches unity. The heralded single photon is subsequently sent to a 50:50 beam splitter (BS), and delocalized over two distinct spatial modes, thus producing the path entangled state \eqref{singlephot}.

To generate the LO in the same time and frequency mode as the heralded photons a second non-linear crystal configured for difference frequency generation (DFG) is employed. For that, the crystal is pumped by the same laser as the HSPS and seeded with a CW laser at 1546\.nm~\cite{Bruno2014b}.  In this configuration, the indistinguishability between the single photon and the coherent state is $>$ 97\%. 
The LO is coupled into a single mode optical fibre, with an orthogonal polarization with respect to that of the path entangled state at the same 50:50 BS used to generate the entangled state. In this way, any phase fluctuations that affect the single photon will equally affect the LO, and the relative phases between the two is maintained even when propagating through fiber. At this stage the coherent states contain roughly 100 photons per pulse. 

Weak displacement measurements are performed  in an all-fiber configuration by Alice, $\cal{\hat{D}(\alpha)}$, and Bob, $\cal{\hat{D}(\beta)}$, mixing their respective share of the entangled state with the LO into a single polarization mode. In the inset of ~\figref{setup} we see the conceptual version of a displacement operation using a variable BS and the equivalent fiber implementation. This is achieved through a set of polarization rotators (PC) followed by a polarizer (LP), which effectively act as variable ratio BS. The polarization rotators consist of three piezo actuators that introduce small pressure-induced birefringence in the optical fibre. The polarizers project part of the LO and the photon onto the same polarization mode, where we can vary both the phase ($\theta$) and amplitude {$r$} of the displacement operations.

The challenge of this experiment is to optimize each element for maximum transmission. The fields in the output modes are finally detected using MoSi superconducting nanowire single-photon detectors (SNSPD)~\cite{Verma2015} with efficiencies of 85\% operating at a temperature of around 1~K. Considering the coupling efficiency of 80\% and the total transmission of all optical elements in the setup of 78\%, we obtain a probability of 52\% to detect the heralded photon in case of no displacement. After a detection the SNSPDs are inactive for a short recovery time, so we placed a pulse picker, based on electro-optic (amplitude) modulation, at the output of the DFG source to reduce the rate of the experiment to 9.5 MHz. We then use a logic gate to only herald entangled states when the coherent state is present, achieving a repetition rate for heralded entanglement of $\sim$2~kHz. To perform the data analysis all the detection events are recorded using a time-to-digital converter. 

The bound $S_{max}$ corresponds to an ideal displacement and perfect single-photon detection. In the experiment, the displacement is implemented using a BS of finite transmittivity, and the single-photon detectors have finite efficiency. Since Bob is a trusted party, this can in principle be accounted for if these parameters are measured, and will lead to a lower value of the bound. However, we use the more stringent bound, which is not influenced by experimental uncertainties on the detector efficiency, i.e. that all the losses due to the detector inefficiency are considered as losses before the displacement. 

\begin{figure}[t!]
\includegraphics[width=0.5\textwidth]{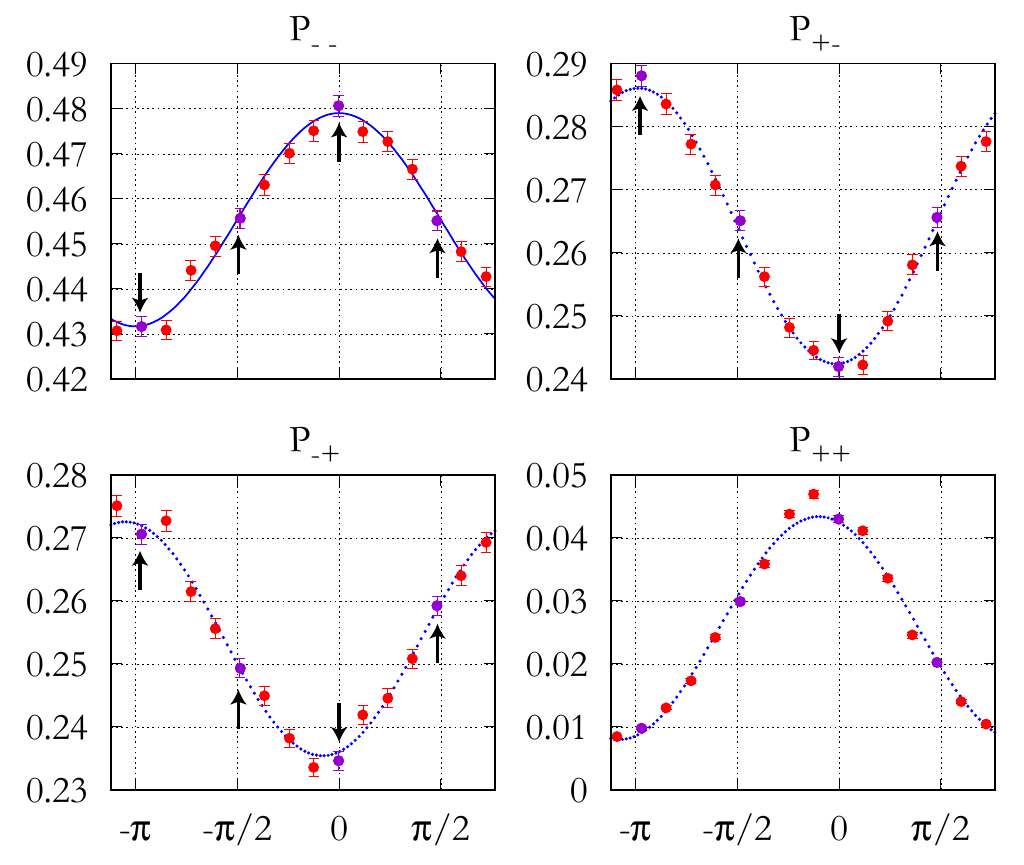}
\caption{\label{results} Observed joint probabilities $p(a,b)$ for click/no-click events of Alice and Bob as functions of the relative phase between displacements (red points). A fit (blue lines) is used to identify which points (arrows) are used to compute the steering value. Error bars (sometimes smaller than the points) correspond to 1 standard deviation.}
\end{figure}

To set the displacement amplitudes, we measure the probability of obtaining a detection when only the coherent state is present and obtain $ r_{A} = 0.233 \pm 0.013$ and $ r_{B} = 0.217 \pm 0.005$, which according to theoretical, modeling should give a clear violation of \eqref{eq.ineqprobs}. In order to implement the different measurement settings, we must vary the phase of the displacements. We implement an active phase change on Alice's side, while Bob's remains fixed. We vary Alice's phase (thus changing the relative phase between Alice and Bob) in small steps and record the number of detection and non-detection events. From the results, we extract the joint probabilities $p(a,b)$ as a function of Alice's phase, shown in \figref{results}. 

To obtain the probabilities $p(a,b|x,y)$, we then pick four points on the curve (indicated by arrows) corresponding to $x=1,2,3,4$ and $y=1$. In order to obtain the probabilities for $y=2,3,4$, we observe that fixing Bob's phase corresponds to choosing a given reference frame. Note that the phase of the LO is not well defined, in other words there is no preferred reference frame. If the amplitude $r_A$ is independent of the phase, then going from one frame to another (i.e. changing Bob's phase) corresponds to a permutation of the labels of Alice's measurements. Here we will assume the latter, which allows us to extract all probabilities $p(a,b|x,y)$ from the data, and hence test the steering inequality. 

This analysis leads to $\Delta S_{exp} = S - S_{max} = (4.95 \pm  1.24)\cdot 10^{-3}$, i.e. a violation of the steering inequality by $4$ standard deviations (details about the error calculation are given in the supplementary information in Appendix). To cross-check this result, we fit the data of \figref{results} to a cosine (as expected from theoretical modeling) and extract $p(a,b|x,y)$ from the fit. We obtain $ \Delta S_{fit} = (2.19 \pm  1.05)\cdot 10^{-3}$. This is in good agreement with theoretical predictions (obtained from the estimated density matrix and experimental parameters): $ \Delta S_{theo} = (2.07 \pm 0.03)\cdot 10^{-3}$. The fact that $\Delta S_{exp}$ gives a larger value is primarily due to the data point corresponding to maximal correlations being slightly above the fit.

{\it Discussion.} We have demonstrated steering of a single-photon entangled state via local weak-displacement measurements based on a novel steering inequality adapted to our setup. The 4 standard deviation violation represents a conclusive measurement of single-photon entanglement in a one-sided DI scenario, with applications to partially DI protocols, such as one-sided DI cryptography.

As our setup is completely free of any post-selection, it is in principle directly amenable to a loophole-free Bell inequality test. This would require increasing the detection efficiency from about $50 \%$ to more than $83.5 \%$ for a bipartite test, or $> 74 \%$ for 4-partite Bell tests with three settings per party. While the bipartite case is clearly challenging~\cite{Shalm2015,*Giustina2015}, a similar setup has already demonstrated 3-partite entanglement~\cite{Monteiro2015}. This represents a promising platform for future implementations of DI and semi-DI protocols at high rates.

{\it Acknowledgments.} The authors would like to thank Natalia Bruno for assistance with the HSPS and LO setup. This work was supported by the Swiss national science foundation (Grant No. 200021\_159592 and Starting Grant DIAQ), the NCCR-QSIT, the OTKA grant K111734, as well as the EU project SIQS. NIST acknowledges funding from the DARPA QUINESS program. Part of the research on the detectors was carried out at the Jet Propulsion Laboratory, California Institute of Technology, under a contract with the National Aeronautics and Space Administration.


\appendix
\section{Steering inequalities for displacement-based measurements}

In this section, we provide a numerical procedure to obtain steering inequalities which are suited for displacement-based measurements, in the 0-1-photon subspace.

We consider the following measurements performed by Alice
\begin{align}
 M'(r,\theta) & = 2\Pi'(r,\theta)-\mathbbm{1} \nonumber \\
 &= 
 \begin{pmatrix}
  2e^{-r^2}-1 & 2r e^{-r^2-i\theta}\\
  2r e^{-r^2+i\theta} & 2r^2 e^{-r^2}-1\\
 \end{pmatrix}
\label{meas}
\end{align}
on the state
\begin{equation}
\rho_{AB}(\eta) = \eta\ket{\Psi}\bra{\Psi} + (1-\eta)\ket{0,0}\bra{0,0}
\label{state}
\end{equation}

\noindent where $\Psi = (\ket{0,1}+\ket{1,0})/\sqrt 2$ is defined by Eq.~(1) in the main text and $\eta$ stands for a parameter characterizing detection efficiency. Below we consider a four-setting ($x=1,2,3,4$) two-output ($a=\pm 1$) steering scenario. Accordingly, we assume that Alice fixes $r=r_A$ and pick the phases $\theta_x$. Alice upon performing measurement $x$ on the state~(\ref{state}) and obtaining result $a$ will steer Bob's state into the assemblage $\{\sigma_{a|x}\}_{a,x}$:
\begin{align}
\sigma_{a|x}(\eta) &= \Tr_A{\left(\rho_{AB}(\eta)\Pi_{a|x}\otimes\openone\right)}\nonumber\\
&= \eta \sigma_{a|x}^S + (1-\eta)\sigma_{a|x}^{US},
\label{assemblage}
\end{align}
where
\begin{align}
\sigma_{a|x}^S &= \Tr_A{\left(\ket{\Psi}\bra{\Psi}\Pi_{a|x}\otimes\openone\right)}\nonumber\\
\sigma_{a|x}^{US} &= \Tr_A{\left(\ket{0,0}\bra{0,0}\Pi_{a|x}\otimes\openone\right)}
\label{SandUS}
\end{align}
and the POVM element $\Pi_{a|x}$ is given by
\begin{equation}
\Pi_{a|x}=\frac{\openone + a\, M'(r_A,\theta_x)}{2} .
\label{Max}
\end{equation}

\begin{figure}[t]
    \includegraphics[trim=1cm 1cm 8cm 4cm, clip=true,width=\columnwidth]{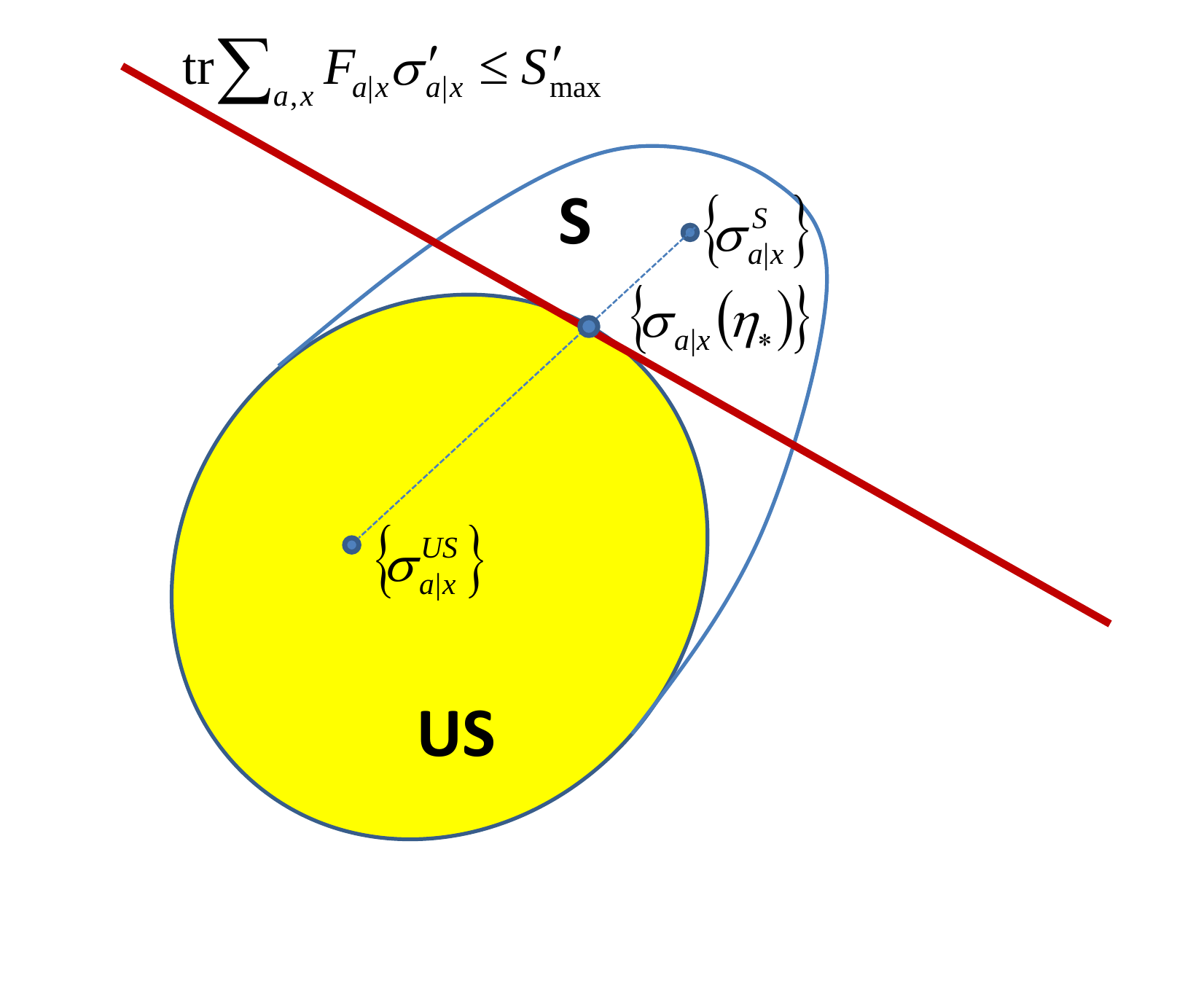} 
    \caption{\label{fig1}
Schematic geometric picture of the different convex sets. US (S) denotes the set of unsteerable (steerable) assemblages. The hyperplane corresponding to a steering inequality is represented by a red line. Any assemblage $\{\sigma_{a|x}'\}_{a,x}$ on the
right-hand side of the red line is detected to be steerable.}
\end{figure}

Our goal now is to solve the semidefinite program (SDP) below based on the works~\cite{Pusey2013, Skrzypczyk2014}

\begin{equation} \label{SDPetacrit}
\begin{aligned}
\eta_{*} \equiv &\text{max} \quad \eta \\
\text{s.t.}\quad &\sum_\lambda D_\lambda(a|x)\sigma_\lambda = \eta\sigma_{a|x}^S(\eta) + (1-\eta)\sigma_{a|x}^{US}(\eta)  &\forall a,x \\
&\Tr \sum_\lambda \sigma_\lambda = 1, \quad \sigma_\lambda \geq 0 &\forall \lambda,
\end{aligned}
\end{equation}

\noindent where the local hidden states $\sigma_{\lambda}$ labeled by $\lambda=(0000,0001,\ldots,1111)$ are the optimization variables, and the matrices $\sigma^S_{a|x}$ and $\sigma^{US}_{a|x}$ are computed in Eq.~(\ref{SandUS}), whereas $D_{\lambda}(a|x)$ define Alice's extremal deterministic strategies. The above SDP returns the critical value $\eta_*$, above which the state $\rho_{AB}(\eta)$ is steerable for a given set of $\theta_x$ and $r_A$ values. Geometrically, the SDP above solves the separation problem sketched in \figref{fig1}.

From the dual solution of the SDP program, it is possible to extract the positive matrices $F_{a|x}$ and the scalar $S'_{\textrm{max}}$, which define the steering inequality, $\Tr \sum_{a,x}{F_{a|x}\sigma'_{a|x}}\le S'_{\textrm{max}}$. Any assemblage $\{\sigma'_{a|x}\}_{a,x}$ providing a larger value than $S'_{\textrm{max}}$ is steerable, and hence certifies that the shared state from which it was obtained is entangled.

Making use of $\sigma'_{-|x}=\sigma'_R-\sigma'_{+|x}$, where $\sigma'_R$ is Bob's reduced state, we arrive at a simpler form of the inequality, as given in the main text 
\begin{equation}
S'\equiv \Tr\left[G'_R\sigma'_R+\sum_{x=1}^4
G'_x\sigma'_{+|x}\right]\le S'_{\textrm{max}} \label{Gineq}
\end{equation}
where $G'_R=\sum_{x=1}^4 F_{-|x}$ and $G'_x=F_{+|x}-F_{-|x}$.

The SDP code~(\ref{SDPetacrit}) computes the critical efficiency $\eta_*$ for fixed parameters of the magnitude $r_A$ and the angles $\theta_x$, $x=1,\ldots,4$ of the displacement operations. We first pick an experimentally relevant value of $r_A=0.2$, and choose some random angles $\theta_x$. Starting from these random angles, a downhill simplex method is used to minimize the value of $\eta_*$ by varying the values of $\theta_x$ for the fixed parameter $r_A=0.2$. This method is a heuristic one, however, independent runs of the above search converge in most of the cases to the set of angles $\theta_x=(0,\pi/2,\pi,3\pi/2)$ and the corresponding set of steering matrices
\begin{equation}
 G'_R =
 \begin{pmatrix}
  s & 0\\
  0 & 0 \\
 \end{pmatrix}
\label{GRmat}
\end{equation}
and
\begin{equation}
 G'_x =
 \begin{pmatrix}
  0 & t e^{\frac{i(x-1)\pi}{4}}\\
  t e^{\frac{-i(x-1)\pi}{4}} & \frac{1}{4}\\
 \end{pmatrix}
\label{Gxmat}
\end{equation}
for $x=1,2,3,4$, parameterized by $s>0$ and $t>0$. The optimal values of the pair $s,t$ providing the lowest $\eta_*$ depend on the value of $r_A$ in general.

We note that the four-setting steering inequality defined by $G'_R$ and $G'_x$ in the ineq.~(\ref{Gineq}) readily generalizes to any number $m\geq4$ of settings as well:
\begin{equation}
 G'_x =
 \begin{pmatrix}
  0 & t e^{\frac{i(x-1)\pi}{m}}\\
  t e^{\frac{-i(x-1)\pi}{m}} & \frac{1}{m}\\
 \end{pmatrix}
\label{Gxmatm}
\end{equation}
for $x=1,2,\ldots,m$ with the same $G'_R$ as in Eq.~(\ref{GRmat}). The unsteerable bound $S'_{\textrm{max}}$ for this $m$-setting inequality is given by the maximum eigenvalue of the matrices $G'_R+\sum_{x=1}^m{l_x G'_x}$ for each choice of $l_{x}=0,1$ similarly to the $m=4$-setting scenario.

\section{Extending the steering inequalities to the full space}

In this section, we provide additional information on how to extend the family of steering inequalities above to the full Fock space, and give the coefficients for the specific inequality used in our experiment.

As explained in the main text, we decompose the matrices $G_R'$, $G_x'$ in terms of the identity operator and a set of measurement operators of the form \eqref{meas} with a fixed $r=r_B$ and some set of angles $\theta_y$, and then extend to the full space by replacing \eqref{meas} by the measurement operator in the full space. We pick the set $\{0,\pi/2,\pi,3\pi/2\}$. This gives an overcomplete set of operators spanning the space of 2x2 matrices. Hence, there is no unique decomposition of the $G'$-matrices, and different decompositions will result in different steering inequalities in the full space that may be harder or easier to violate. We could have picked a smaller set of angles, e.g.~$\{0,2\pi/3,4\pi/3\}$, resulting in a linearly independent set of operators and hence unique decompositions. However, we have found that the following decomposition results in a favourable steering inequality in the full space in terms of ease of violation by our (approximate) single-photon state.

We first decompose each $G'$-matrix on the set $\{\sigma_X,\sigma_Y,\sigma_Z,\mathbbm{1}\}$ consisting of the Pauli operators and the identity. This decomposition is unique. Then we use the following decomposition of the Pauli operators in terms of the measurements $M(r,\theta_y)$
\begin{align}
\sigma_X & = \frac{1}{2 e^{-r^2} r} [ \Pi'(r,0) -  \Pi'(r,\pi) ] \\
\sigma_Y & = \frac{1}{2 e^{-r^2} r} [ \Pi'(r,\frac{\pi}{2}) -  \Pi'(r,\frac{3\pi}{2}) ] \\
\sigma_Z & = \frac{1}{e^{-r^2} (r^2-1)} [ e^{-r^2} (r^2+1) \mathbbm{1} \nonumber\\
& - \Pi'(r,0) - \Pi'(r,\pi) ] .
\end{align}
This results in the following decompositions (c.f.~the main text)
\begin{equation}
\label{eq.Gdecomp}
G_\nu' = \sum_{y=1}^4 c_{\nu y} \, \Pi'(r_B,\theta_y) + c_{\nu 0} \mathbbm{1},
\end{equation}
with
\begin{align}
c_{R0} & = -\frac{r_B^2 s}{1-r_B^2} , \\
c_{R1} = c_{R3} & =  \frac{e^{r_B^2} s}{2 (1-r_B^2)} , \\
c_{R2} = c_{R4} & = 0 ,
\end{align}
\begin{align}
c_{x0} & = \frac{1}{4(1-r_B^2)} \hspace{0.5cm} x = 1,2,3,4 , \\
c_{11} = c_{33} & = \frac{1}{8} e^{r_B^2} \left(\frac{-1}{1-r_B^2}+\frac{4 t}{r_B}\right) , \\
c_{13} = c_{31} & = \frac{1}{8} e^{r_B^2} \left(\frac{-1}{1-r_B^2}-\frac{4 t}{r_B}\right) , \\
c_{12} = c_{14} & = c_{32} = c_{34} = 0 ,
\end{align}
\begin{align}
c_{21} = c_{23} = c_{41} = c_{43} & = \frac{-e^{r_B^2}}{8 \left(1-r_B^2\right)} , \\
c_{22} = c_{44} & = -\frac{e^{r_B^2} t}{2 r_B} , \\
c_{24} = c_{42} & = \frac{e^{r_B^2} t}{2 r_B} .
\end{align}
This is the decomposition we use to define our steering inequality in the full space. That is, we fix the coefficients $c_{\nu y}$, $c_{\nu 0}$ in \eqref{eq.Gdecomp} and replace $\Pi'(r_B,\theta)$ by the full-space operator $\Pi(r_B,\theta) = \ketbra{\alpha}{\alpha}$, where $\alpha = r_Be^{i\theta}$, to get
\begin{equation}
\label{eq.Gfullspace}
G_\nu = \sum_{y=1}^4 c_{\nu y} \, \Pi(r_B,\theta_y) + c_{\nu 0} \mathbbm{1} .
\end{equation}
To get the unsteerable bound in the full space, we need to compute the largest eigenvalue of any of the matrices $G_R+\sum_{x=1}^m{l_x G_x}$ with $l_{x}=0,1$. For given values of $r_B$, $s$, and $t$, we compute these eigenvalues numerically by using the Fock basis representation of $\ket{\alpha} = e^{-r^2/2} \sum_{n=0}^\infty \frac{r^n e^{i\theta n}}{\sqrt{n!}}\ket{n}$ to express the $G$'s as matrices in the Fock basis, and then introducing a cut-off in photon number. We increase the cut-off until the numerically found maximal eigenvalue no longer changes. For small $r_B$, the cut-off does not need to be very large. E.g.~for $r_B=0.2$, a cut-off at $n\leq 4$ is sufficient. 

Finally, to get a steering inequality in terms of the observed probabilities $p(ab|xy)$, we use the relation between Bob's projector and the POVM elements corresponding to the outcomes $b=\pm 1$, which is $\Pi_{+|y} = \Pi(r_B,\theta_y)$ and $\Pi_{-|y} = \mathbbm{1} - \Pi(r_B,\theta_y)$. We also have $\mathbbm{1} = \Pi_{+|y} + \Pi_{-|y}$ for any $y$. In particular, we can take $\mathbbm{1} = \frac{1}{4}\sum_y (\Pi_{+|y} + \Pi_{-|y})$. We insert in \eqref{eq.Gfullspace} and then substitute in the expression for $S$ from the main text
\begin{equation}
S = \tr \left[ G_R\sigma_R + \sum_{x=1}^{4} G_x \sigma_{+|x} \right] .
\end{equation}
We also use that $\sigma_R = \sum_a \sigma_{a|x}$ for any $x$, so in particular we can take $\sigma_R = \frac{1}{4} \sum_{a,x} \sigma_{a|x}$. We then get terms of the form
\begin{equation}
\tr \left[ \Pi_{b|y} \sigma_{a|x} \right] = p(ab|xy) .
\end{equation}
Like this, we obtain an inequality of the form given in the main text
\begin{equation}
S = \sum_{x,y=1}^4 \sum_{a,b=\pm 1} c^{ab}_{xy} p(ab|xy) + c_0 \leq S_{max} ,
\end{equation}
with the new coefficients given by
\begin{align}
c^{++}_{xy} & = c_{xy} + \frac{1}{4} c_{Ry} + \frac{1}{4} c_{x0} , \\
c^{+-}_{xy} & = \frac{1}{4} c_{x0} , \\
c^{-+}_{xy} & = \frac{1}{4} c_{Ry} , \\
c^{--}_{xy} & = 0 , \\
c_0 & = c_{R0} .
\end{align}

\subsubsection{Experimental inequality}

For the experiment, we take $s = 0.983$, $t=0.0656$, and we have $r_B = 0.21 \pm 0.01$. From these numbers we obtain the coefficients (taking the mean value of $r_B$; the effect of the uncertainty on $r_B$ is discussed below)
\begin{equation}
c^{++}_{xy} : \left(
\begin{array}{cccc}
 0.48 & 0.46 & 0.43 & 0.45 \\
 0.46 & 0.43 & 0.45 & 0.48 \\
 0.43 & 0.45 & 0.48 & 0.46 \\
 0.45 & 0.48 & 0.46 & 0.43 \\
\end{array}
\right) ,
\end{equation}
\begin{equation}
c^{+-}_{xy} : \left(
\begin{array}{cccc}
 0.07 & 0.07 & 0.07 & 0.07 \\
 0.07 & 0.07 & 0.07 & 0.07 \\
 0.07 & 0.07 & 0.07 & 0.07 \\
 0.07 & 0.07 & 0.07 & 0.07 \\
\end{array}
\right) ,
\end{equation}
\begin{equation}
c^{-+}_{xy} : \left(
\begin{array}{cccc}
 0.14 & 0 & 0.14 & 0 \\
 0.14 & 0 & 0.14 & 0 \\
 0.14 & 0 & 0.14 & 0 \\
 0.14 & 0 & 0.14 & 0 \\
\end{array}
\right) ,
\end{equation}
\begin{equation}
c_0 = -0.06 .
\end{equation}

\section{Error estimation}

Calculation of the steering value $S$ from experimental data will be affected by uncertainties in the estimated probabilities $p(ab|xy)$, and both $S$ and the bound $S_{max}$ are affected by uncertainty or fluctuations in the value of $r_B$, the amplitude of Bob's displacement. To account for the fact that a perturbation in $r_B$ affects both $S$ and $S_{max}$, we rewrite the steering inequality as $S-S_{max} \leq 0$ and perform the error analysis for the quantity $S-S_{max}$.

\begin{figure}
\includegraphics[width = \columnwidth]{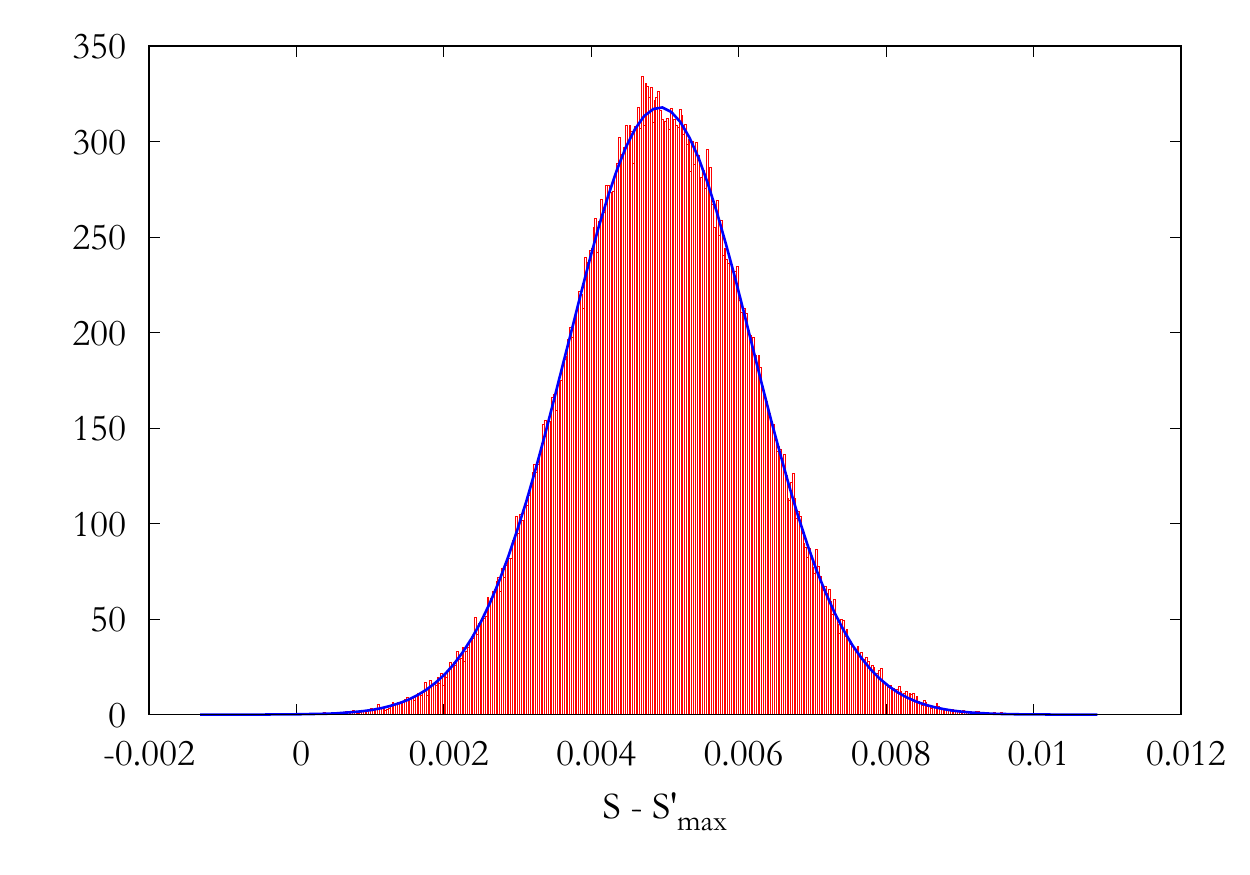}
\caption{\label{fig_histo} Histogram of the simulated value of $S-S_{max}$ based on the experimental uncertainties. The blue curve corresponds to the fit used to estimate the standard deviations.}
\end{figure}

To properly define the error bar with all the error sources, we have used a type of Monte-Carlo algorithm. In each run it randomly draws a number $\tilde N_{xy}^{ab}$ from a poissonian distribution with mean parameter $N_{xy}^{ab}$ which corresponds to the number of measured events with outcome $(a,b)$ for settings $(x,y)$, and it draws $\tilde{r}_B$ from a gaussian distribution with a mean value of 0.217 and standard deviation 0.005. The simulated probabilities $\tilde{p}(ab|xy) = \tilde{N}^{ab}_{xy}/(\sum_{ab} \tilde{N}^{ab}_{xy})$ are then computed and the corresponding value of $S-S_{max}$ is found. After 200000 runs of this procedure we obtained the histogram presented in \figref{fig_histo}, which has a mean value of $4.95 \times 10^{-3}$  with a standard deviation of $1.24\times 10^{-3}$. 

\bibliography{bibliography_steering}

\begin{thebibliography}{41}%
\makeatletter
\providecommand \@ifxundefined [1]{%
 \@ifx{#1\undefined}
}%
\providecommand \@ifnum [1]{%
 \ifnum #1\expandafter \@firstoftwo
 \else \expandafter \@secondoftwo
 \fi
}%
\providecommand \@ifx [1]{%
 \ifx #1\expandafter \@firstoftwo
 \else \expandafter \@secondoftwo
 \fi
}%
\providecommand \natexlab [1]{#1}%
\providecommand \enquote  [1]{``#1''}%
\providecommand \bibnamefont  [1]{#1}%
\providecommand \bibfnamefont [1]{#1}%
\providecommand \citenamefont [1]{#1}%
\providecommand \href@noop [0]{\@secondoftwo}%
\providecommand \href [0]{\begingroup \@sanitize@url \@href}%
\providecommand \@href[1]{\@@startlink{#1}\@@href}%
\providecommand \@@href[1]{\endgroup#1\@@endlink}%
\providecommand \@sanitize@url [0]{\catcode `\\12\catcode `\$12\catcode
  `\&12\catcode `\#12\catcode `\^12\catcode `\_12\catcode `\%12\relax}%
\providecommand \@@startlink[1]{}%
\providecommand \@@endlink[0]{}%
\providecommand \url  [0]{\begingroup\@sanitize@url \@url }%
\providecommand \@url [1]{\endgroup\@href {#1}{\urlprefix }}%
\providecommand \urlprefix  [0]{URL }%
\providecommand \Eprint [0]{\href }%
\providecommand \doibase [0]{http://dx.doi.org/}%
\providecommand \selectlanguage [0]{\@gobble}%
\providecommand \bibinfo  [0]{\@secondoftwo}%
\providecommand \bibfield  [0]{\@secondoftwo}%
\providecommand \translation [1]{[#1]}%
\providecommand \BibitemOpen [0]{}%
\providecommand \bibitemStop [0]{}%
\providecommand \bibitemNoStop [0]{.\EOS\space}%
\providecommand \EOS [0]{\spacefactor3000\relax}%
\providecommand \BibitemShut  [1]{\csname bibitem#1\endcsname}%
\let\auto@bib@innerbib\@empty
\bibitem [{\citenamefont {Tan}\ \emph {et~al.}(1991)\citenamefont {Tan},
  \citenamefont {Walls},\ and\ \citenamefont {Collett}}]{Tan1991}%
  \BibitemOpen
  \bibfield  {author} {\bibinfo {author} {\bibfnamefont {S.~M.}\ \bibnamefont
  {Tan}}, \bibinfo {author} {\bibfnamefont {D.~F.}\ \bibnamefont {Walls}}, \
  and\ \bibinfo {author} {\bibfnamefont {M.~J.}\ \bibnamefont {Collett}},\
  }\href {\doibase 10.1103/PhysRevLett.66.252} {\bibfield  {journal} {\bibinfo
  {journal} {Phys. Rev. Lett.}\ }\textbf {\bibinfo {volume} {66}},\ \bibinfo
  {pages} {252} (\bibinfo {year} {1991})}\BibitemShut {NoStop}%
\bibitem [{\citenamefont {van Enk}(2005)}]{vanEnk2005}%
  \BibitemOpen
  \bibfield  {author} {\bibinfo {author} {\bibfnamefont {S.~J.}\ \bibnamefont
  {van Enk}},\ }\href {\doibase 10.1103/PhysRevA.72.064306} {\bibfield
  {journal} {\bibinfo  {journal} {Phys. Rev. A}\ }\textbf {\bibinfo {volume}
  {72}},\ \bibinfo {pages} {064306} (\bibinfo {year} {2005})}\BibitemShut
  {NoStop}%
\bibitem [{\citenamefont {Sangouard}\ \emph {et~al.}(2011)\citenamefont
  {Sangouard}, \citenamefont {Simon}, \citenamefont {de~Riedmatten},\ and\
  \citenamefont {Gisin}}]{Sangouard2011}%
  \BibitemOpen
  \bibfield  {author} {\bibinfo {author} {\bibfnamefont {N.}~\bibnamefont
  {Sangouard}}, \bibinfo {author} {\bibfnamefont {C.}~\bibnamefont {Simon}},
  \bibinfo {author} {\bibfnamefont {H.}~\bibnamefont {de~Riedmatten}}, \ and\
  \bibinfo {author} {\bibfnamefont {N.}~\bibnamefont {Gisin}},\ }\href
  {\doibase 10.1103/RevModPhys.83.33} {\bibfield  {journal} {\bibinfo
  {journal} {Rev. Mod. Phys.}\ }\textbf {\bibinfo {volume} {83}},\ \bibinfo
  {pages} {33} (\bibinfo {year} {2011})}\BibitemShut {NoStop}%
\bibitem [{\citenamefont {Lombardi}\ \emph {et~al.}(2002)\citenamefont
  {Lombardi}, \citenamefont {Sciarrino}, \citenamefont {Popescu},\ and\
  \citenamefont {De~Martini}}]{Lombardi2002}%
  \BibitemOpen
  \bibfield  {author} {\bibinfo {author} {\bibfnamefont {E.}~\bibnamefont
  {Lombardi}}, \bibinfo {author} {\bibfnamefont {F.}~\bibnamefont {Sciarrino}},
  \bibinfo {author} {\bibfnamefont {S.}~\bibnamefont {Popescu}}, \ and\
  \bibinfo {author} {\bibfnamefont {F.}~\bibnamefont {De~Martini}},\ }\href
  {\doibase 10.1103/PhysRevLett.88.070402} {\bibfield  {journal} {\bibinfo
  {journal} {Phys. Rev. Lett.}\ }\textbf {\bibinfo {volume} {88}},\ \bibinfo
  {pages} {070402} (\bibinfo {year} {2002})}\BibitemShut {NoStop}%
\bibitem [{\citenamefont {Fuwa}\ \emph {et~al.}(2014)\citenamefont {Fuwa},
  \citenamefont {Toba}, \citenamefont {Takeda}, \citenamefont {Marek},
  \citenamefont {Mi\ifmmode~\check{s}\else \v{s}\fi{}ta}, \citenamefont
  {Filip}, \citenamefont {van Loock}, \citenamefont {Yoshikawa},\ and\
  \citenamefont {Furusawa}}]{Fuwa2014}%
  \BibitemOpen
  \bibfield  {author} {\bibinfo {author} {\bibfnamefont {M.}~\bibnamefont
  {Fuwa}}, \bibinfo {author} {\bibfnamefont {S.}~\bibnamefont {Toba}}, \bibinfo
  {author} {\bibfnamefont {S.}~\bibnamefont {Takeda}}, \bibinfo {author}
  {\bibfnamefont {P.}~\bibnamefont {Marek}}, \bibinfo {author} {\bibfnamefont
  {L.}~\bibnamefont {Mi\ifmmode~\check{s}\else \v{s}\fi{}ta}}, \bibinfo
  {author} {\bibfnamefont {R.}~\bibnamefont {Filip}}, \bibinfo {author}
  {\bibfnamefont {P.}~\bibnamefont {van Loock}}, \bibinfo {author}
  {\bibfnamefont {J.-I.}\ \bibnamefont {Yoshikawa}}, \ and\ \bibinfo {author}
  {\bibfnamefont {A.}~\bibnamefont {Furusawa}},\ }\href {\doibase
  10.1103/PhysRevLett.113.223602} {\bibfield  {journal} {\bibinfo  {journal}
  {Phys. Rev. Lett.}\ }\textbf {\bibinfo {volume} {113}},\ \bibinfo {pages}
  {223602} (\bibinfo {year} {2014})}\BibitemShut {NoStop}%
\bibitem [{\citenamefont {Sciarrino}\ \emph {et~al.}(2002)\citenamefont
  {Sciarrino}, \citenamefont {Lombardi}, \citenamefont {Milani},\ and\
  \citenamefont {De~Martini}}]{Sciarrino2002}%
  \BibitemOpen
  \bibfield  {author} {\bibinfo {author} {\bibfnamefont {F.}~\bibnamefont
  {Sciarrino}}, \bibinfo {author} {\bibfnamefont {E.}~\bibnamefont {Lombardi}},
  \bibinfo {author} {\bibfnamefont {G.}~\bibnamefont {Milani}}, \ and\ \bibinfo
  {author} {\bibfnamefont {F.}~\bibnamefont {De~Martini}},\ }\href {\doibase
  10.1103/PhysRevA.66.024309} {\bibfield  {journal} {\bibinfo  {journal} {Phys.
  Rev. A}\ }\textbf {\bibinfo {volume} {66}},\ \bibinfo {pages} {024309}
  (\bibinfo {year} {2002})}\BibitemShut {NoStop}%
\bibitem [{\citenamefont {Osorio}\ \emph {et~al.}(2012)\citenamefont {Osorio},
  \citenamefont {Bruno}, \citenamefont {Sangouard}, \citenamefont {Zbinden},
  \citenamefont {Gisin},\ and\ \citenamefont {Thew}}]{Osorio2012}%
  \BibitemOpen
  \bibfield  {author} {\bibinfo {author} {\bibfnamefont {C.~I.}\ \bibnamefont
  {Osorio}}, \bibinfo {author} {\bibfnamefont {N.}~\bibnamefont {Bruno}},
  \bibinfo {author} {\bibfnamefont {N.}~\bibnamefont {Sangouard}}, \bibinfo
  {author} {\bibfnamefont {H.}~\bibnamefont {Zbinden}}, \bibinfo {author}
  {\bibfnamefont {N.}~\bibnamefont {Gisin}}, \ and\ \bibinfo {author}
  {\bibfnamefont {R.~T.}\ \bibnamefont {Thew}},\ }\href {\doibase
  10.1103/PhysRevA.86.023815} {\bibfield  {journal} {\bibinfo  {journal} {Phys.
  Rev. A}\ }\textbf {\bibinfo {volume} {86}},\ \bibinfo {pages} {023815}
  (\bibinfo {year} {2012})}\BibitemShut {NoStop}%
\bibitem [{\citenamefont {Salart}\ \emph {et~al.}(2010)\citenamefont {Salart},
  \citenamefont {Landry}, \citenamefont {Sangouard}, \citenamefont {Gisin},
  \citenamefont {Herrmann}, \citenamefont {Sanguinetti}, \citenamefont {Simon},
  \citenamefont {Sohler}, \citenamefont {Thew}, \citenamefont {Thomas},\ and\
  \citenamefont {Zbinden}}]{Salart2010}%
  \BibitemOpen
  \bibfield  {author} {\bibinfo {author} {\bibfnamefont {D.}~\bibnamefont
  {Salart}}, \bibinfo {author} {\bibfnamefont {O.}~\bibnamefont {Landry}},
  \bibinfo {author} {\bibfnamefont {N.}~\bibnamefont {Sangouard}}, \bibinfo
  {author} {\bibfnamefont {N.}~\bibnamefont {Gisin}}, \bibinfo {author}
  {\bibfnamefont {H.}~\bibnamefont {Herrmann}}, \bibinfo {author}
  {\bibfnamefont {B.}~\bibnamefont {Sanguinetti}}, \bibinfo {author}
  {\bibfnamefont {C.}~\bibnamefont {Simon}}, \bibinfo {author} {\bibfnamefont
  {W.}~\bibnamefont {Sohler}}, \bibinfo {author} {\bibfnamefont {R.~T.}\
  \bibnamefont {Thew}}, \bibinfo {author} {\bibfnamefont {A.}~\bibnamefont
  {Thomas}}, \ and\ \bibinfo {author} {\bibfnamefont {H.}~\bibnamefont
  {Zbinden}},\ }\href {\doibase 10.1103/PhysRevLett.104.180504} {\bibfield
  {journal} {\bibinfo  {journal} {Phys. Rev. Lett.}\ }\textbf {\bibinfo
  {volume} {104}},\ \bibinfo {pages} {180504} (\bibinfo {year}
  {2010})}\BibitemShut {NoStop}%
\bibitem [{\citenamefont {Papp}\ \emph {et~al.}(2009)\citenamefont {Papp},
  \citenamefont {Choi}, \citenamefont {Deng}, \citenamefont {Lougovski},
  \citenamefont {van Enk},\ and\ \citenamefont {Kimble}}]{Papp2009}%
  \BibitemOpen
  \bibfield  {author} {\bibinfo {author} {\bibfnamefont {S.~B.}\ \bibnamefont
  {Papp}}, \bibinfo {author} {\bibfnamefont {K.~S.}\ \bibnamefont {Choi}},
  \bibinfo {author} {\bibfnamefont {H.}~\bibnamefont {Deng}}, \bibinfo {author}
  {\bibfnamefont {P.}~\bibnamefont {Lougovski}}, \bibinfo {author}
  {\bibfnamefont {S.~J.}\ \bibnamefont {van Enk}}, \ and\ \bibinfo {author}
  {\bibfnamefont {H.~J.}\ \bibnamefont {Kimble}},\ }\href {\doibase
  10.1126/science.1172260} {\bibfield  {journal} {\bibinfo  {journal}
  {Science}\ }\textbf {\bibinfo {volume} {324}},\ \bibinfo {pages} {764}
  (\bibinfo {year} {2009})}\BibitemShut {NoStop}%
\bibitem [{\citenamefont {Gr{\"a}fe}\ \emph {et~al.}(2014)\citenamefont
  {Gr{\"a}fe}, \citenamefont {Heilmann}, \citenamefont {Perez-Leija},
  \citenamefont {Keil}, \citenamefont {Dreisow}, \citenamefont {Heinrich},
  \citenamefont {Moya-Cessa}, \citenamefont {Nolte}, \citenamefont
  {Christodoulides},\ and\ \citenamefont {Szameit}}]{Grafe2014}%
  \BibitemOpen
  \bibfield  {author} {\bibinfo {author} {\bibfnamefont {M.}~\bibnamefont
  {Gr{\"a}fe}}, \bibinfo {author} {\bibfnamefont {R.}~\bibnamefont {Heilmann}},
  \bibinfo {author} {\bibfnamefont {A.}~\bibnamefont {Perez-Leija}}, \bibinfo
  {author} {\bibfnamefont {R.}~\bibnamefont {Keil}}, \bibinfo {author}
  {\bibfnamefont {F.}~\bibnamefont {Dreisow}}, \bibinfo {author} {\bibfnamefont
  {M.}~\bibnamefont {Heinrich}}, \bibinfo {author} {\bibfnamefont
  {H.}~\bibnamefont {Moya-Cessa}}, \bibinfo {author} {\bibfnamefont
  {S.}~\bibnamefont {Nolte}}, \bibinfo {author} {\bibfnamefont {D.~N.}\
  \bibnamefont {Christodoulides}}, \ and\ \bibinfo {author} {\bibfnamefont
  {A.}~\bibnamefont {Szameit}},\ }\href {\doibase 10.1038/nphoton.2014.204}
  {\bibfield  {journal} {\bibinfo  {journal} {Nat Photon}\ }\textbf {\bibinfo
  {volume} {8}},\ \bibinfo {pages} {791} (\bibinfo {year} {2014})}\BibitemShut
  {NoStop}%
\bibitem [{\citenamefont {Ralph}\ and\ \citenamefont {Lund}(2009)}]{Ralph2009}%
  \BibitemOpen
  \bibfield  {author} {\bibinfo {author} {\bibfnamefont {T.~C.}\ \bibnamefont
  {Ralph}}\ and\ \bibinfo {author} {\bibfnamefont {A.~P.}\ \bibnamefont
  {Lund}},\ }\href {\doibase http://dx.doi.org/10.1063/1.3131295} {\bibfield
  {journal} {\bibinfo  {journal} {AIP Conference Proceedings}\ }\textbf
  {\bibinfo {volume} {1110}},\ \bibinfo {pages} {155} (\bibinfo {year}
  {2009})}\BibitemShut {NoStop}%
\bibitem [{\citenamefont {Kocsis}\ \emph {et~al.}(2013)\citenamefont {Kocsis},
  \citenamefont {Xiang}, \citenamefont {Ralph},\ and\ \citenamefont
  {Pryde}}]{Kocsis2013}%
  \BibitemOpen
  \bibfield  {author} {\bibinfo {author} {\bibfnamefont {S.}~\bibnamefont
  {Kocsis}}, \bibinfo {author} {\bibfnamefont {G.~Y.}\ \bibnamefont {Xiang}},
  \bibinfo {author} {\bibfnamefont {T.~C.}\ \bibnamefont {Ralph}}, \ and\
  \bibinfo {author} {\bibfnamefont {G.~J.}\ \bibnamefont {Pryde}},\ }\href
  {\doibase 10.1038/nphys2469} {\bibfield  {journal} {\bibinfo  {journal} {Nat.
  Phys.}\ }\textbf {\bibinfo {volume} {9}},\ \bibinfo {pages} {23} (\bibinfo
  {year} {2013})}\BibitemShut {NoStop}%
\bibitem [{\citenamefont {Bruno}\ \emph {et~al.}(2016)\citenamefont {Bruno},
  \citenamefont {Pini}, \citenamefont {Martin}, \citenamefont {Verma},
  \citenamefont {Nam}, \citenamefont {Mirin}, \citenamefont {Lita},
  \citenamefont {Marsili}, \citenamefont {Korzh}, \citenamefont
  {Bussi{\`{e}}res}, \citenamefont {Sangouard}, \citenamefont {Zbinden},
  \citenamefont {Gisin},\ and\ \citenamefont {Thew}}]{Bruno2016}%
  \BibitemOpen
  \bibfield  {author} {\bibinfo {author} {\bibfnamefont {N.}~\bibnamefont
  {Bruno}}, \bibinfo {author} {\bibfnamefont {V.}~\bibnamefont {Pini}},
  \bibinfo {author} {\bibfnamefont {A.}~\bibnamefont {Martin}}, \bibinfo
  {author} {\bibfnamefont {V.~B.}\ \bibnamefont {Verma}}, \bibinfo {author}
  {\bibfnamefont {S.~W.}\ \bibnamefont {Nam}}, \bibinfo {author} {\bibfnamefont
  {R.}~\bibnamefont {Mirin}}, \bibinfo {author} {\bibfnamefont
  {A.}~\bibnamefont {Lita}}, \bibinfo {author} {\bibfnamefont {F.}~\bibnamefont
  {Marsili}}, \bibinfo {author} {\bibfnamefont {B.}~\bibnamefont {Korzh}},
  \bibinfo {author} {\bibfnamefont {F.}~\bibnamefont {Bussi{\`{e}}res}},
  \bibinfo {author} {\bibfnamefont {N.}~\bibnamefont {Sangouard}}, \bibinfo
  {author} {\bibfnamefont {H.}~\bibnamefont {Zbinden}}, \bibinfo {author}
  {\bibfnamefont {N.}~\bibnamefont {Gisin}}, \ and\ \bibinfo {author}
  {\bibfnamefont {R.~T.}\ \bibnamefont {Thew}},\ }\href {\doibase
  10.1364/OE.24.000125} {\bibfield  {journal} {\bibinfo  {journal} {Opt.
  Express}\ }\textbf {\bibinfo {volume} {24}},\ \bibinfo {pages} {125}
  (\bibinfo {year} {2016})}\BibitemShut {NoStop}%
\bibitem [{\citenamefont {Banaszek}\ and\ \citenamefont
  {W\'odkiewicz}(1999)}]{Banaszek1999}%
  \BibitemOpen
  \bibfield  {author} {\bibinfo {author} {\bibfnamefont {K.}~\bibnamefont
  {Banaszek}}\ and\ \bibinfo {author} {\bibfnamefont {K.}~\bibnamefont
  {W\'odkiewicz}},\ }\href {\doibase 10.1103/PhysRevLett.82.2009} {\bibfield
  {journal} {\bibinfo  {journal} {Phys. Rev. Lett.}\ }\textbf {\bibinfo
  {volume} {82}},\ \bibinfo {pages} {2009} (\bibinfo {year}
  {1999})}\BibitemShut {NoStop}%
\bibitem [{\citenamefont {Bj{\"{o}}rk}\ \emph {et~al.}(2001)\citenamefont
  {Bj{\"{o}}rk}, \citenamefont {Jonsson},\ and\ \citenamefont
  {S{\'{a}}nchez-Soto}}]{Bjork2001}%
  \BibitemOpen
  \bibfield  {author} {\bibinfo {author} {\bibfnamefont {G.}~\bibnamefont
  {Bj{\"{o}}rk}}, \bibinfo {author} {\bibfnamefont {P.}~\bibnamefont
  {Jonsson}}, \ and\ \bibinfo {author} {\bibfnamefont {L.}~\bibnamefont
  {S{\'{a}}nchez-Soto}},\ }\href {\doibase 10.1103/PhysRevA.64.042106}
  {\bibfield  {journal} {\bibinfo  {journal} {Phys. Rev. A}\ }\textbf {\bibinfo
  {volume} {64}},\ \bibinfo {pages} {042106} (\bibinfo {year}
  {2001})}\BibitemShut {NoStop}%
\bibitem [{\citenamefont {Brask}\ and\ \citenamefont
  {Chaves}(2012)}]{Brask2012}%
  \BibitemOpen
  \bibfield  {author} {\bibinfo {author} {\bibfnamefont {J.~B.}\ \bibnamefont
  {Brask}}\ and\ \bibinfo {author} {\bibfnamefont {R.}~\bibnamefont {Chaves}},\
  }\href {\doibase 10.1103/PhysRevA.86.010103} {\bibfield  {journal} {\bibinfo
  {journal} {Phys. Rev. A}\ }\textbf {\bibinfo {volume} {86}},\ \bibinfo
  {pages} {010103} (\bibinfo {year} {2012})}\BibitemShut {NoStop}%
\bibitem [{\citenamefont {Brask}\ \emph {et~al.}(2013)\citenamefont {Brask},
  \citenamefont {Chaves},\ and\ \citenamefont {Brunner}}]{Brask2013}%
  \BibitemOpen
  \bibfield  {author} {\bibinfo {author} {\bibfnamefont {J.~B.}\ \bibnamefont
  {Brask}}, \bibinfo {author} {\bibfnamefont {R.}~\bibnamefont {Chaves}}, \
  and\ \bibinfo {author} {\bibfnamefont {N.}~\bibnamefont {Brunner}},\ }\href
  {\doibase 10.1103/PhysRevA.88.012111} {\bibfield  {journal} {\bibinfo
  {journal} {Phys. Rev. A}\ }\textbf {\bibinfo {volume} {88}},\ \bibinfo
  {pages} {012111} (\bibinfo {year} {2013})}\BibitemShut {NoStop}%
\bibitem [{\citenamefont {Vivoli}\ \emph {et~al.}(2015)\citenamefont {Vivoli},
  \citenamefont {Sekatski}, \citenamefont {Bancal}, \citenamefont {Lim},
  \citenamefont {Martin}, \citenamefont {Thew}, \citenamefont {Zbinden},
  \citenamefont {Gisin},\ and\ \citenamefont {Sangouard}}]{Caprara2015}%
  \BibitemOpen
  \bibfield  {author} {\bibinfo {author} {\bibfnamefont {V.~C.}\ \bibnamefont
  {Vivoli}}, \bibinfo {author} {\bibfnamefont {P.}~\bibnamefont {Sekatski}},
  \bibinfo {author} {\bibfnamefont {J.~D.}\ \bibnamefont {Bancal}}, \bibinfo
  {author} {\bibfnamefont {C.}~\bibnamefont {Lim}}, \bibinfo {author}
  {\bibfnamefont {A.}~\bibnamefont {Martin}}, \bibinfo {author} {\bibfnamefont
  {R.}~\bibnamefont {Thew}}, \bibinfo {author} {\bibfnamefont {H.}~\bibnamefont
  {Zbinden}}, \bibinfo {author} {\bibfnamefont {N.}~\bibnamefont {Gisin}}, \
  and\ \bibinfo {author} {\bibfnamefont {N.}~\bibnamefont {Sangouard}},\ }\href
  {\doibase 10.1088/1367-2630/17/2/023023} {\bibfield  {journal} {\bibinfo
  {journal} {New J. Phys.}\ }\textbf {\bibinfo {volume} {17}},\ \bibinfo
  {pages} {023023} (\bibinfo {year} {2015})}\BibitemShut {NoStop}%
\bibitem [{\citenamefont {Pironio}\ \emph {et~al.}(2010)\citenamefont
  {Pironio}, \citenamefont {Acin}, \citenamefont {Massar}, \citenamefont {de~la
  Giroday}, \citenamefont {Matsukevich}, \citenamefont {Maunz}, \citenamefont
  {Olmschenk}, \citenamefont {Hayes}, \citenamefont {Luo}, \citenamefont
  {Manning},\ and\ \citenamefont {Monroe}}]{Pironio2010}%
  \BibitemOpen
  \bibfield  {author} {\bibinfo {author} {\bibfnamefont {S.}~\bibnamefont
  {Pironio}}, \bibinfo {author} {\bibfnamefont {A.}~\bibnamefont {Acin}},
  \bibinfo {author} {\bibfnamefont {S.}~\bibnamefont {Massar}}, \bibinfo
  {author} {\bibfnamefont {A.~B.}\ \bibnamefont {de~la Giroday}}, \bibinfo
  {author} {\bibfnamefont {D.~N.}\ \bibnamefont {Matsukevich}}, \bibinfo
  {author} {\bibfnamefont {P.}~\bibnamefont {Maunz}}, \bibinfo {author}
  {\bibfnamefont {S.}~\bibnamefont {Olmschenk}}, \bibinfo {author}
  {\bibfnamefont {D.}~\bibnamefont {Hayes}}, \bibinfo {author} {\bibfnamefont
  {L.}~\bibnamefont {Luo}}, \bibinfo {author} {\bibfnamefont {T.~A.}\
  \bibnamefont {Manning}}, \ and\ \bibinfo {author} {\bibfnamefont
  {C.}~\bibnamefont {Monroe}},\ }\href {\doibase 10.1038/nature09008}
  {\bibfield  {journal} {\bibinfo  {journal} {Nature}\ }\textbf {\bibinfo
  {volume} {464}},\ \bibinfo {pages} {1021} (\bibinfo {year}
  {2010})}\BibitemShut {NoStop}%
\bibitem [{\citenamefont {Hofmann}\ \emph {et~al.}(2012)\citenamefont
  {Hofmann}, \citenamefont {Krug}, \citenamefont {Ortegel}, \citenamefont
  {Gerard}, \citenamefont {Weber}, \citenamefont {Rosen},\ and\ \citenamefont
  {Weinfurter}}]{Hofmann2012}%
  \BibitemOpen
  \bibfield  {author} {\bibinfo {author} {\bibfnamefont {J.}~\bibnamefont
  {Hofmann}}, \bibinfo {author} {\bibfnamefont {M.}~\bibnamefont {Krug}},
  \bibinfo {author} {\bibfnamefont {N.}~\bibnamefont {Ortegel}}, \bibinfo
  {author} {\bibfnamefont {L.}~\bibnamefont {Gerard}}, \bibinfo {author}
  {\bibfnamefont {M.}~\bibnamefont {Weber}}, \bibinfo {author} {\bibfnamefont
  {W.}~\bibnamefont {Rosen}}, \ and\ \bibinfo {author} {\bibfnamefont
  {H.}~\bibnamefont {Weinfurter}},\ }\href@noop {} {\bibfield  {journal}
  {\bibinfo  {journal} {Science}\ }\textbf {\bibinfo {volume} {337}},\ \bibinfo
  {pages} {72} (\bibinfo {year} {2012})}\BibitemShut {NoStop}%
\bibitem [{\citenamefont {Ritter}\ \emph {et~al.}(2012)\citenamefont {Ritter},
  \citenamefont {N{\"o}lleke}, \citenamefont {Hahn}, \citenamefont {Reiserer},
  \citenamefont {Neuzner}, \citenamefont {Uphoff}, \citenamefont {M{\"u}cke},
  \citenamefont {Figueroa}, \citenamefont {Bochmann},\ and\ \citenamefont
  {Rempe}}]{Ritter2012}%
  \BibitemOpen
  \bibfield  {author} {\bibinfo {author} {\bibfnamefont {S.}~\bibnamefont
  {Ritter}}, \bibinfo {author} {\bibfnamefont {C.}~\bibnamefont {N{\"o}lleke}},
  \bibinfo {author} {\bibfnamefont {C.}~\bibnamefont {Hahn}}, \bibinfo {author}
  {\bibfnamefont {A.}~\bibnamefont {Reiserer}}, \bibinfo {author}
  {\bibfnamefont {A.}~\bibnamefont {Neuzner}}, \bibinfo {author} {\bibfnamefont
  {M.}~\bibnamefont {Uphoff}}, \bibinfo {author} {\bibfnamefont
  {M.}~\bibnamefont {M{\"u}cke}}, \bibinfo {author} {\bibfnamefont
  {E.}~\bibnamefont {Figueroa}}, \bibinfo {author} {\bibfnamefont
  {J.}~\bibnamefont {Bochmann}}, \ and\ \bibinfo {author} {\bibfnamefont
  {G.}~\bibnamefont {Rempe}},\ }\href@noop {} {\bibfield  {journal} {\bibinfo
  {journal} {Nature}\ }\textbf {\bibinfo {volume} {484}},\ \bibinfo {pages}
  {195} (\bibinfo {year} {2012})}\BibitemShut {NoStop}%
\bibitem [{\citenamefont {Hensen}\ \emph {et~al.}(2015)\citenamefont {Hensen},
  \citenamefont {Bernien}, \citenamefont {Dr{\'e}au}, \citenamefont {Reiserer},
  \citenamefont {Kalb}, \citenamefont {Blok}, \citenamefont {Ruitenberg},
  \citenamefont {Vermeulen}, \citenamefont {Schouten}, \citenamefont
  {Abell{\'a}n}, \citenamefont {Amaya}, \citenamefont {Pruneri}, \citenamefont
  {Mitchell}, \citenamefont {Markham}, \citenamefont {Twitchen}, \citenamefont
  {Elkouss}, \citenamefont {Wehner}, \citenamefont {Taminiau},\ and\
  \citenamefont {Hanson}}]{Hensen2015}%
  \BibitemOpen
  \bibfield  {author} {\bibinfo {author} {\bibfnamefont {B.}~\bibnamefont
  {Hensen}}, \bibinfo {author} {\bibfnamefont {H.}~\bibnamefont {Bernien}},
  \bibinfo {author} {\bibfnamefont {A.}~\bibnamefont {Dr{\'e}au}}, \bibinfo
  {author} {\bibfnamefont {A.}~\bibnamefont {Reiserer}}, \bibinfo {author}
  {\bibfnamefont {N.}~\bibnamefont {Kalb}}, \bibinfo {author} {\bibfnamefont
  {M.}~\bibnamefont {Blok}}, \bibinfo {author} {\bibfnamefont {J.}~\bibnamefont
  {Ruitenberg}}, \bibinfo {author} {\bibfnamefont {R.}~\bibnamefont
  {Vermeulen}}, \bibinfo {author} {\bibfnamefont {R.}~\bibnamefont {Schouten}},
  \bibinfo {author} {\bibfnamefont {C.}~\bibnamefont {Abell{\'a}n}}, \bibinfo
  {author} {\bibfnamefont {W.}~\bibnamefont {Amaya}}, \bibinfo {author}
  {\bibfnamefont {V.}~\bibnamefont {Pruneri}}, \bibinfo {author} {\bibfnamefont
  {M.~W.}\ \bibnamefont {Mitchell}}, \bibinfo {author} {\bibfnamefont
  {M.}~\bibnamefont {Markham}}, \bibinfo {author} {\bibfnamefont
  {D.}~\bibnamefont {Twitchen}}, \bibinfo {author} {\bibfnamefont
  {D.}~\bibnamefont {Elkouss}}, \bibinfo {author} {\bibfnamefont
  {S.}~\bibnamefont {Wehner}}, \bibinfo {author} {\bibfnamefont
  {T.}~\bibnamefont {Taminiau}}, \ and\ \bibinfo {author} {\bibfnamefont
  {R.}~\bibnamefont {Hanson}},\ }\href@noop {} {\bibfield  {journal} {\bibinfo
  {journal} {Nature}\ }\textbf {\bibinfo {volume} {526}},\ \bibinfo {pages}
  {682} (\bibinfo {year} {2015})}\BibitemShut {NoStop}%
\bibitem [{\citenamefont {Monteiro}\ \emph {et~al.}(2015)\citenamefont
  {Monteiro}, \citenamefont {Caprara~Vivoli}, \citenamefont {Guerreiro},
  \citenamefont {Martin}, \citenamefont {Bancal}, \citenamefont {Zbinden},
  \citenamefont {Thew},\ and\ \citenamefont {Sangouard}}]{Monteiro2015}%
  \BibitemOpen
  \bibfield  {author} {\bibinfo {author} {\bibfnamefont {F.}~\bibnamefont
  {Monteiro}}, \bibinfo {author} {\bibfnamefont {V.}~\bibnamefont
  {Caprara~Vivoli}}, \bibinfo {author} {\bibfnamefont {T.}~\bibnamefont
  {Guerreiro}}, \bibinfo {author} {\bibfnamefont {A.}~\bibnamefont {Martin}},
  \bibinfo {author} {\bibfnamefont {J.-D.}\ \bibnamefont {Bancal}}, \bibinfo
  {author} {\bibfnamefont {H.}~\bibnamefont {Zbinden}}, \bibinfo {author}
  {\bibfnamefont {R.~T.}\ \bibnamefont {Thew}}, \ and\ \bibinfo {author}
  {\bibfnamefont {N.}~\bibnamefont {Sangouard}},\ }\href {\doibase
  10.1103/PhysRevLett.114.170504} {\bibfield  {journal} {\bibinfo  {journal}
  {Phys. Rev. Lett.}\ }\textbf {\bibinfo {volume} {114}},\ \bibinfo {pages}
  {170504} (\bibinfo {year} {2015})}\BibitemShut {NoStop}%
\bibitem [{\citenamefont {Wiseman}\ \emph {et~al.}(2007)\citenamefont
  {Wiseman}, \citenamefont {Jones},\ and\ \citenamefont
  {Doherty}}]{Wiseman2007}%
  \BibitemOpen
  \bibfield  {author} {\bibinfo {author} {\bibfnamefont {H.~M.}\ \bibnamefont
  {Wiseman}}, \bibinfo {author} {\bibfnamefont {S.~J.}\ \bibnamefont {Jones}},
  \ and\ \bibinfo {author} {\bibfnamefont {A.~C.}\ \bibnamefont {Doherty}},\
  }\href {\doibase 10.1103/PhysRevLett.98.140402} {\bibfield  {journal}
  {\bibinfo  {journal} {Phys. Rev. Lett.}\ }\textbf {\bibinfo {volume} {98}},\
  \bibinfo {pages} {140402} (\bibinfo {year} {2007})}\BibitemShut {NoStop}%
\bibitem [{\citenamefont {Wittmann}\ \emph {et~al.}(2012)\citenamefont
  {Wittmann}, \citenamefont {Ramelow}, \citenamefont {Steinlechner},
  \citenamefont {Langford}, \citenamefont {Brunner}, \citenamefont {Wiseman},
  \citenamefont {Ursin},\ and\ \citenamefont {Zeilinger}}]{Wittmann2012}%
  \BibitemOpen
  \bibfield  {author} {\bibinfo {author} {\bibfnamefont {B.}~\bibnamefont
  {Wittmann}}, \bibinfo {author} {\bibfnamefont {S.}~\bibnamefont {Ramelow}},
  \bibinfo {author} {\bibfnamefont {F.}~\bibnamefont {Steinlechner}}, \bibinfo
  {author} {\bibfnamefont {N.~K.}\ \bibnamefont {Langford}}, \bibinfo {author}
  {\bibfnamefont {N.}~\bibnamefont {Brunner}}, \bibinfo {author} {\bibfnamefont
  {H.~M.}\ \bibnamefont {Wiseman}}, \bibinfo {author} {\bibfnamefont
  {R.}~\bibnamefont {Ursin}}, \ and\ \bibinfo {author} {\bibfnamefont
  {A.}~\bibnamefont {Zeilinger}},\ }\href {\doibase
  10.1088/1367-2630/14/5/053030} {\bibfield  {journal} {\bibinfo  {journal}
  {New J. Phys.}\ }\textbf {\bibinfo {volume} {14}},\ \bibinfo {pages} {053030}
  (\bibinfo {year} {2012})}\BibitemShut {NoStop}%
\bibitem [{\citenamefont {Smith}\ \emph {et~al.}(2012)\citenamefont {Smith},
  \citenamefont {Gillett}, \citenamefont {de~Almeida}, \citenamefont
  {Branciard}, \citenamefont {Fedrizzi}, \citenamefont {Weinhold},
  \citenamefont {Lita}, \citenamefont {Calkins}, \citenamefont {Gerrits},
  \citenamefont {Wiseman}, \citenamefont {Nam},\ and\ \citenamefont
  {White}}]{Smith2012}%
  \BibitemOpen
  \bibfield  {author} {\bibinfo {author} {\bibfnamefont {D.~H.}\ \bibnamefont
  {Smith}}, \bibinfo {author} {\bibfnamefont {G.}~\bibnamefont {Gillett}},
  \bibinfo {author} {\bibfnamefont {M.~P.}\ \bibnamefont {de~Almeida}},
  \bibinfo {author} {\bibfnamefont {C.}~\bibnamefont {Branciard}}, \bibinfo
  {author} {\bibfnamefont {A.}~\bibnamefont {Fedrizzi}}, \bibinfo {author}
  {\bibfnamefont {T.~J.}\ \bibnamefont {Weinhold}}, \bibinfo {author}
  {\bibfnamefont {A.}~\bibnamefont {Lita}}, \bibinfo {author} {\bibfnamefont
  {B.}~\bibnamefont {Calkins}}, \bibinfo {author} {\bibfnamefont
  {T.}~\bibnamefont {Gerrits}}, \bibinfo {author} {\bibfnamefont {H.~M.}\
  \bibnamefont {Wiseman}}, \bibinfo {author} {\bibfnamefont {S.~W.}\
  \bibnamefont {Nam}}, \ and\ \bibinfo {author} {\bibfnamefont {A.~G.}\
  \bibnamefont {White}},\ }\href {\doibase 10.1038/ncomms1628} {\bibfield
  {journal} {\bibinfo  {journal} {Nat. Commun.}\ }\textbf {\bibinfo {volume}
  {3}},\ \bibinfo {pages} {625} (\bibinfo {year} {2012})}\BibitemShut {NoStop}%
\bibitem [{\citenamefont {Bennet}\ \emph {et~al.}(2012)\citenamefont {Bennet},
  \citenamefont {Evans}, \citenamefont {Saunders}, \citenamefont {Branciard},
  \citenamefont {Cavalcanti}, \citenamefont {Wiseman},\ and\ \citenamefont
  {Pryde}}]{Bennet2012}%
  \BibitemOpen
  \bibfield  {author} {\bibinfo {author} {\bibfnamefont {A.~J.}\ \bibnamefont
  {Bennet}}, \bibinfo {author} {\bibfnamefont {D.~A.}\ \bibnamefont {Evans}},
  \bibinfo {author} {\bibfnamefont {D.~J.}\ \bibnamefont {Saunders}}, \bibinfo
  {author} {\bibfnamefont {C.}~\bibnamefont {Branciard}}, \bibinfo {author}
  {\bibfnamefont {E.~G.}\ \bibnamefont {Cavalcanti}}, \bibinfo {author}
  {\bibfnamefont {H.~M.}\ \bibnamefont {Wiseman}}, \ and\ \bibinfo {author}
  {\bibfnamefont {G.~J.}\ \bibnamefont {Pryde}},\ }\href {\doibase
  10.1103/PhysRevX.2.031003} {\bibfield  {journal} {\bibinfo  {journal} {Phys.
  Rev. X}\ }\textbf {\bibinfo {volume} {2}},\ \bibinfo {pages} {031003}
  (\bibinfo {year} {2012})}\BibitemShut {NoStop}%
\bibitem [{\citenamefont {Fuwa}\ \emph {et~al.}(2015)\citenamefont {Fuwa},
  \citenamefont {Takeda}, \citenamefont {Zwierz}, \citenamefont {Wiseman},\
  and\ \citenamefont {Furusawa}}]{Fuwa2015}%
  \BibitemOpen
  \bibfield  {author} {\bibinfo {author} {\bibfnamefont {M.}~\bibnamefont
  {Fuwa}}, \bibinfo {author} {\bibfnamefont {S.}~\bibnamefont {Takeda}},
  \bibinfo {author} {\bibfnamefont {M.}~\bibnamefont {Zwierz}}, \bibinfo
  {author} {\bibfnamefont {H.~M.}\ \bibnamefont {Wiseman}}, \ and\ \bibinfo
  {author} {\bibfnamefont {A.}~\bibnamefont {Furusawa}},\ }\href {\doibase
  10.1038/ncomms7665} {\bibfield  {journal} {\bibinfo  {journal} {Nat.
  Commun.}\ }\textbf {\bibinfo {volume} {6}},\ \bibinfo {pages} {6665}
  (\bibinfo {year} {2015})}\BibitemShut {NoStop}%
\bibitem [{\citenamefont {Branciard}\ \emph {et~al.}(2012)\citenamefont
  {Branciard}, \citenamefont {Cavalcanti}, \citenamefont {Walborn},
  \citenamefont {Scarani},\ and\ \citenamefont {Wiseman}}]{Branciard2012}%
  \BibitemOpen
  \bibfield  {author} {\bibinfo {author} {\bibfnamefont {C.}~\bibnamefont
  {Branciard}}, \bibinfo {author} {\bibfnamefont {E.~G.}\ \bibnamefont
  {Cavalcanti}}, \bibinfo {author} {\bibfnamefont {S.~P.}\ \bibnamefont
  {Walborn}}, \bibinfo {author} {\bibfnamefont {V.}~\bibnamefont {Scarani}}, \
  and\ \bibinfo {author} {\bibfnamefont {H.~M.}\ \bibnamefont {Wiseman}},\
  }\href {\doibase 10.1103/PhysRevA.85.010301} {\bibfield  {journal} {\bibinfo
  {journal} {Phys. Rev. A}\ }\textbf {\bibinfo {volume} {85}},\ \bibinfo
  {pages} {010301} (\bibinfo {year} {2012})}\BibitemShut {NoStop}%
\bibitem [{\citenamefont {Cavalcanti}\ \emph {et~al.}(2009)\citenamefont
  {Cavalcanti}, \citenamefont {Jones}, \citenamefont {Wiseman},\ and\
  \citenamefont {Reid}}]{Cavalcanti2009}%
  \BibitemOpen
  \bibfield  {author} {\bibinfo {author} {\bibfnamefont {E.~G.}\ \bibnamefont
  {Cavalcanti}}, \bibinfo {author} {\bibfnamefont {S.~J.}\ \bibnamefont
  {Jones}}, \bibinfo {author} {\bibfnamefont {H.~M.}\ \bibnamefont {Wiseman}},
  \ and\ \bibinfo {author} {\bibfnamefont {M.~D.}\ \bibnamefont {Reid}},\
  }\href {\doibase 10.1103/PhysRevA.80.032112} {\bibfield  {journal} {\bibinfo
  {journal} {Phys. Rev. A}\ }\textbf {\bibinfo {volume} {80}},\ \bibinfo
  {pages} {032112} (\bibinfo {year} {2009})}\BibitemShut {NoStop}%
\bibitem [{\citenamefont {Brunner}\ \emph {et~al.}(2014)\citenamefont
  {Brunner}, \citenamefont {Cavalcanti}, \citenamefont {Pironio}, \citenamefont
  {Scarani},\ and\ \citenamefont {Wehner}}]{Brunner2014}%
  \BibitemOpen
  \bibfield  {author} {\bibinfo {author} {\bibfnamefont {N.}~\bibnamefont
  {Brunner}}, \bibinfo {author} {\bibfnamefont {D.}~\bibnamefont {Cavalcanti}},
  \bibinfo {author} {\bibfnamefont {S.}~\bibnamefont {Pironio}}, \bibinfo
  {author} {\bibfnamefont {V.}~\bibnamefont {Scarani}}, \ and\ \bibinfo
  {author} {\bibfnamefont {S.}~\bibnamefont {Wehner}},\ }\href {\doibase
  10.1103/RevModPhys.86.419} {\bibfield  {journal} {\bibinfo  {journal} {Rev.
  Mod. Phys.}\ }\textbf {\bibinfo {volume} {86}},\ \bibinfo {pages} {419}
  (\bibinfo {year} {2014})}\BibitemShut {NoStop}%
\bibitem [{\citenamefont {Ac\'in}\ \emph {et~al.}(2007)\citenamefont {Ac\'in},
  \citenamefont {Brunner}, \citenamefont {Gisin}, \citenamefont {Massar},
  \citenamefont {Pironio},\ and\ \citenamefont {Scarani}}]{Acin2007}%
  \BibitemOpen
  \bibfield  {author} {\bibinfo {author} {\bibfnamefont {A.}~\bibnamefont
  {Ac\'in}}, \bibinfo {author} {\bibfnamefont {N.}~\bibnamefont {Brunner}},
  \bibinfo {author} {\bibfnamefont {N.}~\bibnamefont {Gisin}}, \bibinfo
  {author} {\bibfnamefont {S.}~\bibnamefont {Massar}}, \bibinfo {author}
  {\bibfnamefont {S.}~\bibnamefont {Pironio}}, \ and\ \bibinfo {author}
  {\bibfnamefont {V.}~\bibnamefont {Scarani}},\ }\href {\doibase
  10.1103/PhysRevLett.98.230501} {\bibfield  {journal} {\bibinfo  {journal}
  {Phys. Rev. Lett.}\ }\textbf {\bibinfo {volume} {98}},\ \bibinfo {pages}
  {230501} (\bibinfo {year} {2007})}\BibitemShut {NoStop}%
\bibitem [{\citenamefont {Paris}(1996)}]{Paris1996}%
  \BibitemOpen
  \bibfield  {author} {\bibinfo {author} {\bibfnamefont {M.~G.~A.}\
  \bibnamefont {Paris}},\ }\href {\doibase 10.1016/0375-9601(96)00339-8}
  {\bibfield  {journal} {\bibinfo  {journal} {Phys. Lett. A}\ }\textbf
  {\bibinfo {volume} {217}},\ \bibinfo {pages} {78} (\bibinfo {year}
  {1996})}\BibitemShut {NoStop}%
\bibitem [{\citenamefont {Pusey}(2013)}]{Pusey2013}%
  \BibitemOpen
  \bibfield  {author} {\bibinfo {author} {\bibfnamefont {M.~F.}\ \bibnamefont
  {Pusey}},\ }\href {\doibase 10.1103/PhysRevA.88.032313} {\bibfield  {journal}
  {\bibinfo  {journal} {Phys. Rev. A}\ }\textbf {\bibinfo {volume} {88}},\
  \bibinfo {pages} {032313} (\bibinfo {year} {2013})}\BibitemShut {NoStop}%
\bibitem [{\citenamefont {Skrzypczyk}\ \emph {et~al.}(2014)\citenamefont
  {Skrzypczyk}, \citenamefont {Navascu\'es},\ and\ \citenamefont
  {Cavalcanti}}]{Skrzypczyk2014}%
  \BibitemOpen
  \bibfield  {author} {\bibinfo {author} {\bibfnamefont {P.}~\bibnamefont
  {Skrzypczyk}}, \bibinfo {author} {\bibfnamefont {M.}~\bibnamefont
  {Navascu\'es}}, \ and\ \bibinfo {author} {\bibfnamefont {D.}~\bibnamefont
  {Cavalcanti}},\ }\href {\doibase 10.1103/PhysRevLett.112.180404} {\bibfield
  {journal} {\bibinfo  {journal} {Phys. Rev. Lett.}\ }\textbf {\bibinfo
  {volume} {112}},\ \bibinfo {pages} {180404} (\bibinfo {year}
  {2014})}\BibitemShut {NoStop}%
\bibitem [{\citenamefont {Bruno}\ \emph
  {et~al.}(2014{\natexlab{a}})\citenamefont {Bruno}, \citenamefont {Martin},
  \citenamefont {Guerreiro}, \citenamefont {Sanguinetti},\ and\ \citenamefont
  {Thew}}]{Bruno2014}%
  \BibitemOpen
  \bibfield  {author} {\bibinfo {author} {\bibfnamefont {N.}~\bibnamefont
  {Bruno}}, \bibinfo {author} {\bibfnamefont {A.}~\bibnamefont {Martin}},
  \bibinfo {author} {\bibfnamefont {T.}~\bibnamefont {Guerreiro}}, \bibinfo
  {author} {\bibfnamefont {B.}~\bibnamefont {Sanguinetti}}, \ and\ \bibinfo
  {author} {\bibfnamefont {R.~T.}\ \bibnamefont {Thew}},\ }\href {\doibase
  10.1364/OE.22.017246} {\bibfield  {journal} {\bibinfo  {journal} {Opt.
  Express}\ }\textbf {\bibinfo {volume} {22}},\ \bibinfo {pages} {17246}
  (\bibinfo {year} {2014}{\natexlab{a}})}\BibitemShut {NoStop}%
\bibitem [{\citenamefont {Guerreiro}\ \emph {et~al.}(2013)\citenamefont
  {Guerreiro}, \citenamefont {Martin}, \citenamefont {Sanguinetti},
  \citenamefont {Bruno}, \citenamefont {Zbinden},\ and\ \citenamefont
  {Thew}}]{Guerreiro2013}%
  \BibitemOpen
  \bibfield  {author} {\bibinfo {author} {\bibfnamefont {T.}~\bibnamefont
  {Guerreiro}}, \bibinfo {author} {\bibfnamefont {A.}~\bibnamefont {Martin}},
  \bibinfo {author} {\bibfnamefont {B.}~\bibnamefont {Sanguinetti}}, \bibinfo
  {author} {\bibfnamefont {N.}~\bibnamefont {Bruno}}, \bibinfo {author}
  {\bibfnamefont {H.}~\bibnamefont {Zbinden}}, \ and\ \bibinfo {author}
  {\bibfnamefont {R.~T.}\ \bibnamefont {Thew}},\ }\href@noop {} {\bibfield
  {journal} {\bibinfo  {journal} {Opt. Express}\ }\textbf {\bibinfo {volume}
  {21}},\ \bibinfo {pages} {27641} (\bibinfo {year} {2013})}\BibitemShut
  {NoStop}%
\bibitem [{\citenamefont {Bruno}\ \emph
  {et~al.}(2014{\natexlab{b}})\citenamefont {Bruno}, \citenamefont {Martin},\
  and\ \citenamefont {Thew}}]{Bruno2014b}%
  \BibitemOpen
  \bibfield  {author} {\bibinfo {author} {\bibfnamefont {N.}~\bibnamefont
  {Bruno}}, \bibinfo {author} {\bibfnamefont {A.}~\bibnamefont {Martin}}, \
  and\ \bibinfo {author} {\bibfnamefont {R.}~\bibnamefont {Thew}},\ }\href
  {\doibase 10.1016/j.optcom.2014.02.025} {\bibfield  {journal} {\bibinfo
  {journal} {Opt. Commun.}\ }\textbf {\bibinfo {volume} {327}},\ \bibinfo
  {pages} {17} (\bibinfo {year} {2014}{\natexlab{b}})}\BibitemShut {NoStop}%
\bibitem [{\citenamefont {Verma}\ \emph {et~al.}(2015)\citenamefont {Verma},
  \citenamefont {Korzh}, \citenamefont {Bussi{\`{e}}res}, \citenamefont
  {Horansky}, \citenamefont {Dyer}, \citenamefont {Lita}, \citenamefont
  {Vayshenker}, \citenamefont {Marsili}, \citenamefont {Shaw}, \citenamefont
  {Zbinden}, \citenamefont {Mirin},\ and\ \citenamefont {Nam}}]{Verma2015}%
  \BibitemOpen
  \bibfield  {author} {\bibinfo {author} {\bibfnamefont {V.~B.}\ \bibnamefont
  {Verma}}, \bibinfo {author} {\bibfnamefont {B.}~\bibnamefont {Korzh}},
  \bibinfo {author} {\bibfnamefont {F.}~\bibnamefont {Bussi{\`{e}}res}},
  \bibinfo {author} {\bibfnamefont {R.~D.}\ \bibnamefont {Horansky}}, \bibinfo
  {author} {\bibfnamefont {S.~D.}\ \bibnamefont {Dyer}}, \bibinfo {author}
  {\bibfnamefont {A.~E.}\ \bibnamefont {Lita}}, \bibinfo {author}
  {\bibfnamefont {I.}~\bibnamefont {Vayshenker}}, \bibinfo {author}
  {\bibfnamefont {F.}~\bibnamefont {Marsili}}, \bibinfo {author} {\bibfnamefont
  {M.~D.}\ \bibnamefont {Shaw}}, \bibinfo {author} {\bibfnamefont
  {H.}~\bibnamefont {Zbinden}}, \bibinfo {author} {\bibfnamefont {R.~P.}\
  \bibnamefont {Mirin}}, \ and\ \bibinfo {author} {\bibfnamefont {S.~W.}\
  \bibnamefont {Nam}},\ }\href {\doibase 10.1364/OE.23.033792} {\bibfield
  {journal} {\bibinfo  {journal} {Opt. Express}\ }\textbf {\bibinfo {volume}
  {23}},\ \bibinfo {pages} {33792} (\bibinfo {year} {2015})}\BibitemShut
  {NoStop}%
\bibitem [{\citenamefont {Shalm}\ \emph {et~al.}(2015)\citenamefont {Shalm},
  \citenamefont {Meyer-Scott}, \citenamefont {Christensen}, \citenamefont
  {Bierhorst}, \citenamefont {Wayne}, \citenamefont {Stevens}, \citenamefont
  {Gerrits}, \citenamefont {Glancy}, \citenamefont {Hamel}, \citenamefont
  {Allman}, \citenamefont {Coakley}, \citenamefont {Dyer}, \citenamefont
  {Hodge}, \citenamefont {Lita}, \citenamefont {Verma}, \citenamefont
  {Lambrocco}, \citenamefont {Tortorici}, \citenamefont {Migdall},
  \citenamefont {Zhang}, \citenamefont {Kumor}, \citenamefont {Farr},
  \citenamefont {Marsili}, \citenamefont {Shaw}, \citenamefont {Stern},
  \citenamefont {Abell\'an}, \citenamefont {Amaya}, \citenamefont {Pruneri},
  \citenamefont {Jennewein}, \citenamefont {Mitchell}, \citenamefont {Kwiat},
  \citenamefont {Bienfang}, \citenamefont {Mirin}, \citenamefont {Knill},\ and\
  \citenamefont {Nam}}]{Shalm2015}%
  \BibitemOpen
  \bibfield  {author} {\bibinfo {author} {\bibfnamefont {L.~K.}\ \bibnamefont
  {Shalm}}, \bibinfo {author} {\bibfnamefont {E.}~\bibnamefont {Meyer-Scott}},
  \bibinfo {author} {\bibfnamefont {B.~G.}\ \bibnamefont {Christensen}},
  \bibinfo {author} {\bibfnamefont {P.}~\bibnamefont {Bierhorst}}, \bibinfo
  {author} {\bibfnamefont {M.~A.}\ \bibnamefont {Wayne}}, \bibinfo {author}
  {\bibfnamefont {M.~J.}\ \bibnamefont {Stevens}}, \bibinfo {author}
  {\bibfnamefont {T.}~\bibnamefont {Gerrits}}, \bibinfo {author} {\bibfnamefont
  {S.}~\bibnamefont {Glancy}}, \bibinfo {author} {\bibfnamefont {D.~R.}\
  \bibnamefont {Hamel}}, \bibinfo {author} {\bibfnamefont {M.~S.}\ \bibnamefont
  {Allman}}, \bibinfo {author} {\bibfnamefont {K.~J.}\ \bibnamefont {Coakley}},
  \bibinfo {author} {\bibfnamefont {S.~D.}\ \bibnamefont {Dyer}}, \bibinfo
  {author} {\bibfnamefont {C.}~\bibnamefont {Hodge}}, \bibinfo {author}
  {\bibfnamefont {A.~E.}\ \bibnamefont {Lita}}, \bibinfo {author}
  {\bibfnamefont {V.~B.}\ \bibnamefont {Verma}}, \bibinfo {author}
  {\bibfnamefont {C.}~\bibnamefont {Lambrocco}}, \bibinfo {author}
  {\bibfnamefont {E.}~\bibnamefont {Tortorici}}, \bibinfo {author}
  {\bibfnamefont {A.~L.}\ \bibnamefont {Migdall}}, \bibinfo {author}
  {\bibfnamefont {Y.}~\bibnamefont {Zhang}}, \bibinfo {author} {\bibfnamefont
  {D.~R.}\ \bibnamefont {Kumor}}, \bibinfo {author} {\bibfnamefont {W.~H.}\
  \bibnamefont {Farr}}, \bibinfo {author} {\bibfnamefont {F.}~\bibnamefont
  {Marsili}}, \bibinfo {author} {\bibfnamefont {M.~D.}\ \bibnamefont {Shaw}},
  \bibinfo {author} {\bibfnamefont {J.~A.}\ \bibnamefont {Stern}}, \bibinfo
  {author} {\bibfnamefont {C.}~\bibnamefont {Abell\'an}}, \bibinfo {author}
  {\bibfnamefont {W.}~\bibnamefont {Amaya}}, \bibinfo {author} {\bibfnamefont
  {V.}~\bibnamefont {Pruneri}}, \bibinfo {author} {\bibfnamefont
  {T.}~\bibnamefont {Jennewein}}, \bibinfo {author} {\bibfnamefont {M.~W.}\
  \bibnamefont {Mitchell}}, \bibinfo {author} {\bibfnamefont {P.~G.}\
  \bibnamefont {Kwiat}}, \bibinfo {author} {\bibfnamefont {J.~C.}\ \bibnamefont
  {Bienfang}}, \bibinfo {author} {\bibfnamefont {R.~P.}\ \bibnamefont {Mirin}},
  \bibinfo {author} {\bibfnamefont {E.}~\bibnamefont {Knill}}, \ and\ \bibinfo
  {author} {\bibfnamefont {S.~W.}\ \bibnamefont {Nam}},\ }\href {\doibase
  10.1103/PhysRevLett.115.250402} {\bibfield  {journal} {\bibinfo  {journal}
  {Phys. Rev. Lett.}\ }\textbf {\bibinfo {volume} {115}},\ \bibinfo {pages}
  {250402} (\bibinfo {year} {2015})}\BibitemShut {NoStop}%
\bibitem [{\citenamefont {Giustina}\ \emph {et~al.}(2015)\citenamefont
  {Giustina}, \citenamefont {Versteegh}, \citenamefont {Wengerowsky},
  \citenamefont {Handsteiner}, \citenamefont {Hochrainer}, \citenamefont
  {Phelan}, \citenamefont {Steinlechner}, \citenamefont {Kofler}, \citenamefont
  {Larsson}, \citenamefont {Abell\'an}, \citenamefont {Amaya}, \citenamefont
  {Pruneri}, \citenamefont {Mitchell}, \citenamefont {Beyer}, \citenamefont
  {Gerrits}, \citenamefont {Lita}, \citenamefont {Shalm}, \citenamefont {Nam},
  \citenamefont {Scheidl}, \citenamefont {Ursin}, \citenamefont {Wittmann},\
  and\ \citenamefont {Zeilinger}}]{Giustina2015}%
  \BibitemOpen
  \bibfield  {author} {\bibinfo {author} {\bibfnamefont {M.}~\bibnamefont
  {Giustina}}, \bibinfo {author} {\bibfnamefont {M.~A.~M.}\ \bibnamefont
  {Versteegh}}, \bibinfo {author} {\bibfnamefont {S.}~\bibnamefont
  {Wengerowsky}}, \bibinfo {author} {\bibfnamefont {J.}~\bibnamefont
  {Handsteiner}}, \bibinfo {author} {\bibfnamefont {A.}~\bibnamefont
  {Hochrainer}}, \bibinfo {author} {\bibfnamefont {K.}~\bibnamefont {Phelan}},
  \bibinfo {author} {\bibfnamefont {F.}~\bibnamefont {Steinlechner}}, \bibinfo
  {author} {\bibfnamefont {J.}~\bibnamefont {Kofler}}, \bibinfo {author}
  {\bibfnamefont {J.-A.}\ \bibnamefont {Larsson}}, \bibinfo {author}
  {\bibfnamefont {C.}~\bibnamefont {Abell\'an}}, \bibinfo {author}
  {\bibfnamefont {W.}~\bibnamefont {Amaya}}, \bibinfo {author} {\bibfnamefont
  {V.}~\bibnamefont {Pruneri}}, \bibinfo {author} {\bibfnamefont {M.~W.}\
  \bibnamefont {Mitchell}}, \bibinfo {author} {\bibfnamefont {J.}~\bibnamefont
  {Beyer}}, \bibinfo {author} {\bibfnamefont {T.}~\bibnamefont {Gerrits}},
  \bibinfo {author} {\bibfnamefont {A.~E.}\ \bibnamefont {Lita}}, \bibinfo
  {author} {\bibfnamefont {L.~K.}\ \bibnamefont {Shalm}}, \bibinfo {author}
  {\bibfnamefont {S.~W.}\ \bibnamefont {Nam}}, \bibinfo {author} {\bibfnamefont
  {T.}~\bibnamefont {Scheidl}}, \bibinfo {author} {\bibfnamefont
  {R.}~\bibnamefont {Ursin}}, \bibinfo {author} {\bibfnamefont
  {B.}~\bibnamefont {Wittmann}}, \ and\ \bibinfo {author} {\bibfnamefont
  {A.}~\bibnamefont {Zeilinger}},\ }\href {\doibase
  10.1103/PhysRevLett.115.250401} {\bibfield  {journal} {\bibinfo  {journal}
  {Phys. Rev. Lett.}\ }\textbf {\bibinfo {volume} {115}},\ \bibinfo {pages}
  {250401} (\bibinfo {year} {2015})}\BibitemShut {NoStop}%
\end{thebibliography}%


%

\end{document}